\documentclass[a4paper,hidelinks, 12pt]{article}
\usepackage[utf8]{inputenc}
\usepackage{a4wide}
\usepackage{amsmath}
\usepackage{amsfonts}
\usepackage{amssymb}
\usepackage{amsthm}

\usepackage{subfigure}
\usepackage{cite}
\usepackage{xcolor}
\usepackage{soul}
\usepackage{cancel}
\usepackage[toc,page]{appendix}
\numberwithin{equation}{section}
\usepackage{mathtools}
\usepackage{hyperref}
\usepackage{enumitem}

\allowdisplaybreaks

\newcommand\ztwo{\mathbb{Z}_2}
\def\Prms#1#2{\left(P_{#1#2}\right)_\text{r.m.s}}

\newcommand{\half}{{\textstyle\frac{1}{2}}}

\renewcommand{\Re}{\mbox{Re\thinspace}}
\renewcommand{\Im}{\mbox{Im\thinspace}}
\def\gsim{\mathrel{\raise.3ex\hbox{$>$\kern-.75em\lower1ex\hbox{$\sim$}}}}
\newcommand{\HB}{{\text{HB}}}

\newtheorem*{con-non}{Conjecture}
\theoremstyle{plain}

\begin{document}
\begin{titlepage}
\begin{center}

{\large \bf {Weinberg's 3HDM potential with spontaneous CP violation}}

\vskip 1cm

R. Plantey,$^{a,}$\footnote{E-mail: Robin.Plantey@ntnu.no} 
O. M. Ogreid,$^{b,}$\footnote{E-mail: omo@hvl.no}
P. Osland,$^{c,}$\footnote{E-mail: Per.Osland@uib.no}
M. N. Rebelo$^{d,}$\footnote{E-mail: rebelo@tecnico.ulisboa.pt} and
M. Aa. Solberg$^{a,}$\footnote{E-mail: Marius.Solberg@ntnu.no} 

\vspace{1.0cm}

$^{a}$Department of Structural Engineering, NTNU, \\
7491 Trondheim, Norway, \\
$^{b}$Western Norway University of Applied Sciences,\\ Postboks 7030, N-5020 Bergen, 
Norway, \\
$^{c}$Department of Physics and Technology, University of Bergen, \\
Postboks 7803, N-5020  Bergen, Norway,\\
$^{d}$Centro de F\'isica Te\'orica de Part\'iculas -- CFTP and Dept de F\' \i sica\\
Instituto Superior T\'ecnico -- IST, Universidade de Lisboa, Av. Rovisco Pais, \\
P-1049-001 Lisboa, Portugal \\
\end{center}

\vskip 3cm

\begin{abstract}
We study the potential of Weinberg's $\mathbb{Z}_2\times\mathbb{Z}_2$-symmetric three-Higgs-doublet model (3HDM). The potential is designed to accommodate CP violation in the scalar sector within a gauge theory, while at the same time allowing for natural flavour conservation. This framework allows for both explicit and spontaneous CP violation. CP can be explicitly violated when the coefficients of the potential are taken to be complex. With coefficients chosen to be real, CP can be spontaneously violated via complex vacuum expectation values (vevs). In the absence of the terms leading to the possibility of CP violation, either explicit or induced by complex vevs, the potential has two global U(1) symmetries. In this case, spontaneous symmetry breaking would in general give rise to massless states. In a realistic implementation, those terms must be included, thus preventing the existence of Goldstone bosons.  A scan over parameters, imposing the existence of a neutral state at 125~GeV that is nearly CP-even shows that, in the absence of fine-tuning, the scalar spectrum contains one or two states with masses below 125~GeV that have a significant CP-odd component. These light states would have a low production rate via the Bjorken process and could thus have escaped detection at LEP. At the LHC the situation is less clear. While we do not here aim for a full phenomenological study of the light states, 
we point out that the $\gamma\gamma$ decay channel would be challenging to measure because of suppressed couplings to $WW$.

\end{abstract}

\end{titlepage}

\setcounter{footnote}{0}

%%%%%%%%%%%%%%%%%%%%%%%%%%%%%%%%%%%%%%%%%%%%%%%%
\section{Introduction}
%%%%%%%%%%%%%%%%%%%%%%%%%%%%%%%%%%%%%%%%%%%%%%%%
In the Standard Model (SM) there is only one Higgs doublet and CP cannot be violated in the scalar sector. With the addition of one extra Higgs doublet CP can be violated in this sector both explicitly, via the introduction of complex coefficients, or spontaneously as was shown by T. D. Lee \cite{Lee:1973iz}. Spontaneous CP violation puts 
the breaking of CP and electroweak symmetry breaking on equal footing. However, the Yukawa couplings of models with two or more Higgs doublets lead to potentially dangerous flavour-changing neutral currents (FCNC), for which there are stringent experimental limits. In order to solve this problem for the two-Higgs-doublet model a solution was proposed \cite{Glashow:1976nt,Paschos:1976ay}, based on the imposition of natural flavour conservation (NFC) resulting from an additional $\mathbb{Z}_2$ symmetry in the scalar and in the Yukawa sector, forcing all the 
right-handed quarks of each sector only to couple to a single Higgs doublet, thus eliminating FCNC at the tree level. However, imposing a discrete symmetry on the scalar potential in the context of two-Higgs-doublet models automatically leads to CP conservation. This can be evaded by adding a term softly breaking the $\mathbb{Z}_2$ symmetry, in which case CP can be spontaneously violated \cite{Branco:1985aq}.
In 1976 it was pointed out by Weinberg \cite{Weinberg:1976hu} that the scalar potential of models with three Higgs doublets and with additional $\mathbb{Z}_2$ symmetries leading to NFC 
can violate CP explicitly
and can also provide a mechanism for naturally small CP violation.  Soon afterwards 
Branco \cite{Branco:1980sz}
showed that this framework also allows for the possibility of spontaneous CP violation. 

 In this work we outline some important features of the Weinberg potential with real coefficients and CP violation, with an emphasis on the mass spectrum. A more detailed phenomenological analysis will be presented elsewhere.

It is also important to point out that the requirements of spontaneous CP breaking and NFC lead to a class of theories where CP non-conservation is solely due to Higgs exchange \cite{Branco:1979pv}. The fact that the right-handed quarks of each sector only couple to a single Higgs doublet allows for the rephasing of the right-handed quarks in such a way as to cancel  the phase of the vev of the doublet to which these quarks couple, thus leading to a real  Cabibbo-Kobayashi-Maskawa (CKM) matrix. It is by now experimentally established that the CKM matrix is complex \cite{Botella:2005fc,Charles:2004jd} implying 
that if one wants to build a fully realistic model from the point of view of flavour this issue must be addressed. To solve this problem one might for instance consider scenarios with the addition of vector-like quarks \cite{Aguilar-Saavedra:2013qpa,Alves:2023ufm}.

We consider the explicitly CP-conserving $\ztwo\times \ztwo$-symmetric\footnote{The potential is separately symmetric under $\phi_i\to-\phi_i$ for all three $\phi_i$, which means that there are in fact three $ \mathbb{Z}_2$ symmetries.} Weinberg potential \cite{Weinberg:1976hu}, following the notation of Ivanov and Nishi \cite{Ivanov:2014doa}
\begin{subequations} \label{Eq:W_pot}
\begin{equation}
  V=V_2+V_4, \quad \text{with}\quad
  V_4=V_0+V_\text{ph},
\end{equation}
where $V_2$ and $V_0$ are insensitive to independent rephasing of the Higgs doublets,
\begin{align} 
  V_2&=-[m_{11}(\phi_1^\dagger \phi_1)+m_{22}(\phi_2^\dagger \phi_2)+m_{33}(\phi_3^\dagger \phi_3)], \\
  V_0&=\lambda_{11}(\phi_1^\dagger\phi_1)^2+\lambda_{12}(\phi_1^\dagger\phi_1)(\phi_2^\dagger\phi_2)
  +\lambda_{13}(\phi_1^\dagger\phi_1)(\phi_3^\dagger\phi_3)+\lambda_{22}(\phi_2^\dagger\phi_2)^2 \nonumber \\
  &+\lambda_{23}(\phi_2^\dagger\phi_2)(\phi_3^\dagger\phi_3)+\lambda_{33}(\phi_3^\dagger\phi_3)^2 \nonumber \\
  &+\lambda^\prime_{12}(\phi_1^\dagger\phi_2)(\phi_2^\dagger\phi_1)
  +\lambda^\prime_{13}(\phi_1^\dagger\phi_3)(\phi_3^\dagger\phi_1)
  +\lambda^\prime_{23}(\phi_2^\dagger\phi_3)(\phi_3^\dagger\phi_2), \label{Eq:V_symm}
\end{align}
whereas
\begin{equation} \label{Eq:V_ph}
V_\text{ph}=\lambda_1(\phi_2^\dagger\phi_3)^2+\lambda_2(\phi_3^\dagger\phi_1)^2 
+\lambda_3(\phi_1^\dagger\phi_2)^2 + \text{ h.c.}
\end{equation}
\end{subequations}
would be sensitive to rephasing of the doublets. Explicit CP conservation means that it is possible to make $\lambda_1$, $\lambda_2$, $\lambda_3$ real by a rephasing of the scalar doublets. In this case CP violation can only occur spontaneously, i.e., via complex vevs. For simplicity, in our discussion we choose to work in this basis.

In the limit of $\{\lambda_1,\lambda_2,\lambda_3\}\to0$ (or $V_\text{ph}\to0$), the potential acquires two\footnote{The third U(1) symmetry can be absorbed in the U(1) hypercharge symmetry.} U(1) symmetries, since both $V_2$ and $V_0$ are insensitive to rephasing of the fields. It is the emergence of an additional symmetry that would allow for these terms to be removed from the potential in a consistent way. Different symmetries of multi-Higgs models occur frequently and play an important role. As is clear from the classification in ref.~\cite{Darvishi:2019dbh}, the full additional symmetry in this limit is simply the U(1)$\times$U(1) symmetry we are seeing here. Starting from the general Weinberg potential, two of the scalar masses tend to zero when we approach the limit where these U(1) global symmetries emerge and are broken by the vacuum.\footnote{The masses are continuous functions of the couplings of the phase-sensitive part of the potential: The masses squared are the roots of the characteristic polynomial of the mass-squared matrix. The coefficients of this characteristic polynomial will be polynomials in the couplings of the phase-sensitive part of the potential, i.e.~continuous functions of these couplings. Moreover, the roots of a polynomial are continuous functions of the coefficients (see e.g.~\cite{Harris:1987}), so then the masses squared are continuous functions of the phase-sensitive couplings.}  

Experimentally, an SM-like scalar ($h_\text{SM}$) has been observed at 125.25~GeV with trilinear $h_\text{SM}VV$ ($V=W,Z$) gauge couplings that have very little CP-odd ``contamination'' \cite{CMS:2019jdw,ATLAS:2020evk}. One way to arrive at this situation is for the coefficients of the phase-sensitive terms of the potential to be small.
In the limit when these terms vanish, CP is conserved and the physical scalars have definite CP parities. As stated earlier, there will also be two massless states in this limit, as long as all vevs are non-zero. 

At this point, it is useful to comment on ``natural'' alignment, when the $h_\text{SM}VV$ coupling automatically attains full strength due to the symmetry of the potential. Pilaftsis has shown \cite{Pilaftsis:2016erj} (see also Ref.~\cite{BhupalDev:2014bir}) that this happens in a 3HDM if the quartic part of the potential has an Sp(6), SU(3) or $\text{SO(3)}\times\text{CP}$ symmetry. Another possibility is to have an unbroken $\ztwo\times \ztwo$ symmetry. In our framework we require CP to be broken spontaneously. In order to have CP violation $\lambda_1$, $\lambda_2$ and $\lambda_3$ must be simultaneously non-zero and all vevs must be different from zero. The latter breaks the $\ztwo\times \ztwo$ symmetry. Therefore there is no natural alignment in this case.
Since both the Weinberg $\ztwo\times \ztwo$ symmetric potential and the $\text{U(1)}\times\text{U(1)}$ symmetric limit contain terms not compatible with these higher symmetries, it follows that natural alignment is not available in the present framework. In particular, we note that CP violation is not compatible with natural alignment.

In this work, we instead enforce alignment as a constraint on the parameters, leaving room for small deviations.

In view of the above discussion, it is interesting to explore whether the spectrum will contain two light states, lighter than the one whose trilinear $hVV$ gauge coupling is SM-like. What we will see in our parameter scans is the following feature:

\textit{In a realistic case, i.e.~with an SM-like Higgs boson at $m_h=125.25$~\text{GeV}, the scenario where the SM-like scalar is the lightest requires fine-tuning. That is, in the bulk of the acceptable parameter space, lighter neutral scalars are predicted. These generally have a considerable CP-odd content.}

Moreover, those light states would have suppressed trilinear gauge couplings $h_iWW$ and $h_iZZ$ ($i=1,2$), since these couplings are constrained by the orthogonality of the mixing matrix. Hence they may have escaped detection at LEP.

The paper is organised as follows. In section~\ref{sect:general} we minimize the Weinberg potential, discuss CP conservation and properties of the mass matrices, introducing at the same time notation and definitions used in the remainder of the article. Section~\ref{sect:gauge} presents the couplings among the electroweak gauge bosons and the scalars, and section~\ref{sect:Yukawa} the Yukawa couplings. Then, in section~\ref{sect:scan} we present results of a scan over the potential parameters, subject to a set of well established constraints. In section~\ref{sect:h_SM} we compare two ways of accommodating the discovered SM-like Higgs particle in this potential, with either one or two states being lighter. Finally, section~\ref{sect:conclusions} contains concluding remarks. The expressions for the mass-squared matrices and pseudo-Goldstone masses are given in appendix~\ref{sect:mass-matrix} and a simple version of the model, which turns out to conserve CP is discussed in Appendix~\ref{sect:app-minimal}.

%%%%%%%%%%%%%%%%%%%%%%%%%%%%%%%%%%%%%%%%%%%%%%%%%
\section{General properties of the Weinberg potential}
\label{sect:general}
%%%%%%%%%%%%%%%%%%%%%%%%%%%%%%%%%%%%%%%%%%%%%%%%%
We give here some basic properties of the minimum of the potential and
comment on conditions for CP conservation. Such conditions can be
analyzed from the point of view of CP-odd scalar basis invariants
\cite{Botella:1994cs,Branco:2005em}  (see also Ref.~\cite{Gunion:2005ja}), but a complete discussion is beyond the scope of this work and will be presented elsewhere. We will here only note that CP is conserved whenever any coupling in $V_\text{ph}$ vanishes (provided all vevs are non-zero) or $\sin(2\theta_2-2\theta_3)=0$.\footnote{Let the indices $\{i,j,k\}$ be some permutation of $\{1,2,3\}$, and consider the vanishing of $\lambda_i$: The minimization conditions will then enforce the vanishing of $\lambda_j$ and $\lambda_k$, unless the angles take on special values. Whenever all $\lambda_l$'s vanish $V_\text{ph}$ also vanishes and all vevs can be made real.}

%%%%%%%%%%%%%%%%%%%%%%%%%%%%%%%%%%%%%%%%%%%%%%%%%
\subsection{Minimizing the potential}
%%%%%%%%%%%%%%%%%%%%%%%%%%%%%%%%%%%%%%%%%%%%%%%%%

By an overall phase rotation, we choose the vev of $\phi_1$, $w_1\equiv v_1$ real, whereas the other vevs, $w_2$ and $w_3$ will in general be complex. We introduce phases $\theta_i$ by
\begin{equation}
w_i=v_i\,e^{i\theta_i}, \quad i=2,3,
\end{equation}
with $v_1^2+v_2^2+v_3^2=v^2$ and $v=246$~GeV.
We will thus represent the different vacua in the form
\begin{equation}
\{w_1,w_2,w_3\}=\{v_1,v_2\,e^{i\theta_2},v_3\,e^{i\theta_3}\}.
\end{equation}

It is convenient to extract an overall phase factor and decompose the SU(2) doublets as
\begin{equation} \label{Eq:phi_i-phase}
\phi_i=e^{i\theta_i}\left(
\begin{array}{c}\phi_i^+\\ (v_i+\eta_i+i \chi_i)/\sqrt{2}
\end{array}\right), \quad i=1,2,3.
\end{equation}
In our convention, $\theta_1=0$, $\phi_1$ being a reference for the phases of the other fields.

In general, CP is violated, so we cannot assign CP parities to the fields $\eta_i$ and $\chi_i$. However, since they are independent fields, they have opposite ``CP content'' in the sense that the product $\eta_i\chi_i$ is odd under CP.

The minimization with respect to the moduli of the vevs gives
\begin{subequations} \label{Eq:eqs-m_ii}
\begin{align}
m_{11}&=\lambda_{11}v_1^2+\half\bar\lambda_{12}v_2^2
+\half\bar\lambda_{13}v_3^2
+\lambda_2\cos(2\theta_3)v_3^2+\lambda_3\cos(2\theta_2)v_2^2, \\
m_{22}&=\lambda_{22}v_2^2+\half\bar\lambda_{12}v_1^2
+\half\bar\lambda_{23}v_3^2
+\lambda_1\cos{(2\theta_3-2\theta_2)}v_3^2+\lambda_3\cos(2\theta_2)v_1^2, \\
m_{33}&=\lambda_{33}v_3^2+\half\bar\lambda_{13}v_1^2
+\half\bar\lambda_{23}v_2^2
+\lambda_1\cos{(2\theta_3-2\theta_2)}v_2^2+\lambda_2\cos(2\theta_3)v_1^2,
\end{align}
\end{subequations}
where we introduced the abbreviations
\begin{equation}
\label{Eq:lambda_bar}
\bar\lambda_{12}\equiv\lambda_{12}+\lambda^\prime_{12}, \quad
\bar\lambda_{13}\equiv\lambda_{13}+\lambda^\prime_{13}, \quad
\bar\lambda_{23}\equiv\lambda_{23}+\lambda^\prime_{23}.
\end{equation}
These abbreviations are also useful for the neutral-sector mass matrices.

There are two minimization constraints with respect to the phases. These can be expressed as
\begin{subequations} \label{Eq:angles}
\begin{align} 
&\lambda_1v_3^2\sin(2\theta_2-2\theta_3)+\lambda_3v_1^2\sin2\theta_2=0, \\
&\lambda_1v_2^2\sin(2\theta_3-2\theta_2)+\lambda_2v_1^2\sin2\theta_3=0.
\end{align}
\end{subequations}
From these two relations, it follows that the two phases are related via
\begin{equation} \label{Eq:constraint-th2-th3}
\lambda_3v_2^2\sin2\theta_2+\lambda_2v_3^2\sin2\theta_3=0.
\end{equation}
It also follows that the relative sign of $\sin2\theta_2$ and $\sin2\theta_3$ is the opposite of the relative sign between $\lambda_2$ and $\lambda_3$.\footnote{The ranges of these parameters could accordingly be reduced.}

One can impose these two conditions (\ref{Eq:angles}) by substituting for $\lambda_2$ and $\lambda_3$:
\begin{subequations} \label{Eq:lam_23}
\begin{align} 
\lambda_2&=\phantom{-}\frac{\lambda_1v_2^2\sin(2\theta_2-2\theta_3)}{v_1^2\sin2\theta_3}, \\
\lambda_3&=-\frac{\lambda_1v_3^2\sin(2\theta_2-2\theta_3)}{v_1^2\sin2\theta_2}.
\end{align}
\end{subequations}
Insisting on perturbativity, we require all $\lambda_i\in[-4\pi,4\pi]$. Thus, whenever $\theta_2$ or $\theta_3$ is small, the other angle must be close (modulo $\pi/2$).

Alternatively, the minimization conditions (\ref{Eq:angles}) yield the solutions \cite{Branco:1980sz}\footnote{These expressions differ from those of Ref.~\cite{Branco:1980sz} since we take $\phi_1$ rather than $\phi_2$ to have a real vev.}
\begin{subequations} \label{Eq:solutions-theta}
\begin{align}
\cos2\theta_2&=\frac{1}{2}\left[\frac{D_{23}D_{31}}{D_{12}^2}-\frac{D_{31}}{D_{23}}-\frac{D_{23}}{D_{31}} \right], \\
\cos2\theta_3&=\frac{1}{2}\left[\frac{D_{23}D_{12}}{D_{31}^2}-\frac{D_{12}}{D_{23}}-\frac{D_{23}}{D_{12}} \right],
\end{align}
\end{subequations}
with
\begin{equation}
D_{12}=\lambda_3(v_1v_2)^2, \quad D_{23}=\lambda_1(v_2v_3)^2, \quad D_{31}=\lambda_2(v_3v_1)^2.
\end{equation}
Interpreting the $D_{ij}$ as sides in a triangle \cite{Branco:1980sz} requires $\lambda_1$, $\lambda_2$ and $\lambda_3$ to all be positive. As noted above, $\theta_2$ and $\theta_3$ must then have opposite signs.

%%%%%%%%%%%%%%%%%%%%%%%%%%%%%%%%%%%%%%%%%%%%%%%%%
\subsection{The case $\theta_2=\theta_3+n\pi/2$}
\label{sect:theta_2=theta_3}
%%%%%%%%%%%%%%%%%%%%%%%%%%%%%%%%%%%%%%%%%%%%%%%%%
When $\theta_2$ and $\theta_3$ differ by a multiple of $\pi/2$, the first terms of Eqs.~(\ref{Eq:angles}) vanish. These minimization conditions then require one of the following to be satisfied (assuming all vevs are non-zero):
\begin{enumerate}
\item
$\lambda_2=\lambda_3=0$,
\item
$\lambda_2=0$, $\sin2\theta_2=0$,
\item
$\lambda_3=0$, $\sin2\theta_3=0$,
\item
$\sin2\theta_2=\sin2\theta_3=0$.
\end{enumerate}
All these cases are CP conserving, and will not be considered in the following.

\paragraph{$\theta_2=\theta_3$:}
When $\theta_2=\theta_3$ we may go to a basis in which $w_2$ and $w_3$ are real, and $w_1$ is complex. It then follows that we have only one minimization condition with respect to phases, there will remain a ``left-over'' field on which the mass-squared matrix does not depend, i.e., a massless state.

\paragraph{$\theta_2=\theta_3\pm\pi$:}
This case is essentially equivalent to the case above, except for some sign changes.

\paragraph{$\theta_2=\theta_3\pm\pi/2$:}
This case is also essentially equivalent to the case above, except for an interchange of the $\eta_i$ and $\chi_i$ fields in one doublet.

%%%%%%%%%%%%%%%%%%%%%%%%%%%%%%%%%%%%%%%%%%%%%%%%%
\subsection{Rotating to a Higgs basis}
\label{sect:HB}
%%%%%%%%%%%%%%%%%%%%%%%%%%%%%%%%%%%%%%%%%%%%%%%%%
To make these mass-squared matrices as simple as possible, and to easily identify the SM Higgs 
in the neutral mass spectrum (cf.\ Eq.~\eqref{E:OSM} below), it is convenient to rotate the Higgs doublets to a Higgs basis, where only one doublet has a non-zero vev.

A suitable Higgs basis is reached by the transformation
\begin{equation}
 {\cal R}_2 {\cal R}_1 
\begin{pmatrix} v_1 \\ e^{i\theta_2} v_2 \\ e^{i\theta_3} v_3 \end{pmatrix}
=\begin{pmatrix} v \\ 0 \\ 0 \end{pmatrix}.
\end{equation}
with
\begin{equation}
{\cal R}_1=
\begin{pmatrix}
1 & 0  \\
0 & R_1
\end{pmatrix}, \quad 
R_1=\frac{1}{w}
\begin{pmatrix}
v_2 e^{-i\theta_2} & v_3e^{-i\theta_3} \\
-v_3e^{-i\theta_2} & v_2e^{-i\theta_3}
\end{pmatrix}, \quad
w=\sqrt{v_2^2+v_3^2},
\end{equation}
and
\begin{equation}
{\cal R}_2=\frac{1}{v}
\begin{pmatrix}
v_1 & w & 0 \\
-w & v_1 & 0 \\
0 & 0 & v
\end{pmatrix}.
\end{equation}

Thus, the Higgs basis (with SU(2) doublets $H_1$, $H_2$ and $H_3$) is reached by ${\cal R}\equiv  {\cal R}_2 {\cal R}_1 $,
\begin{equation} \label{Eq:toHiggsBasis}
\begin{pmatrix}
H_1 \\ H_2 \\ H_3
\end{pmatrix}
={\cal R}
\begin{pmatrix}
\phi_1 \\ \phi_2 \\ \phi_3
\end{pmatrix}
=\tilde{\cal R}
\begin{pmatrix}
\phi_1 \\ e^{-i\theta_2}\phi_2 \\ e^{-i\theta_3}\phi_3
\end{pmatrix},
\end{equation}
with
\begin{equation} \label{eq:Rtilde}
\tilde{\cal R}={\cal R}_2\frac{1}{w}
\begin{pmatrix}
w & 0 & 0\\
0 & v_2 & v_3 \\
0 & -v_3 & v_2
\end{pmatrix}
\end{equation}
in fact real.

We decompose the Higgs-basis fields as
\begin{equation} \label{Eq:cap_H_i}
H_1=\left(
\begin{array}{c}G^+\\ (v+\eta^\text{HB}_1+i G_0)/\sqrt{2}
\end{array}\right), \quad 
H_i=\left(
\begin{array}{c}\varphi_{i}^\text{HB +}\\ (\eta^\text{HB}_i+i \chi^\text{HB}_i)/\sqrt{2}
\end{array}\right), \quad 
i=2,3,
\end{equation}
and enumerate the neutral fields \{1,2,3,4,5\} according to the following sequence:
\begin{equation} \label{Eq:field-sequence}
\varphi_i^\text{HB }=\{\eta_1^\HB, \quad \eta_2^\HB, \quad \eta_3^\HB, \quad \chi_2^\HB, \quad \chi_3^\HB\}, \quad i=1,\ldots 5.
\end{equation}
%%%%%%%%%%%%%%%%%%%%%%%%%%%%%%%%%%%%%%%%%%%%%%%%%
\subsection{Masses}
\label{sect:masses}
%%%%%%%%%%%%%%%%%%%%%%%%%%%%%%%%%%%%%%%%%%%%%%%%%
The elements of the $2\times 2$ charged mass-squared matrix ${\cal M}^2_\text{ch}$, as well as the masses squared, are given in Appendix \ref{sect:charged-sector}, while the elements of the
$5 \times 5$ neutral mass-squared matrix ${\cal M}^2_\text{neut}$ are given in 
Appendix \ref{sect:neutral-sector}. Moreover, we give $\mathcal O(\lambda_1)$ formulas for the masses squared of the pseudo-Goldstone bosons in Appendix \ref{sect:low-lam1}.

We diagonalize the general neutral mass-squared matrix by a $5\times5$ rotation matrix $O$ to obtain the mass eigenstates:
\begin{equation} \label{Eq:Orthogonal}
h_i=O_{ij}\varphi_j^\text{HB},
\end{equation}
with $\varphi_j^\text{HB}$ defined by Eq.~(\ref{Eq:field-sequence}).

Since the mass-squared matrix of the neutral sector is $5\times5$, the rotation matrix $O$ of Eq.~(\ref{Eq:Orthogonal}) can only be numerically determined. This somewhat limits our analysis.
In Appendix \ref{sect:neutral-sector} we schematically quote the determinant (\ref{Eq:det-5by5}) of the neutral-sector mass-squared matrix. It is proportional to $\lambda_1^2$, reflecting the fact that the potential has two massless states in the limit $\lambda_1\to0$.

In Appendix~\ref{sect:app-minimal} we briefly discuss a ``minimal'' version of the potential, with $\lambda_3=\pm\lambda_2$, $\theta_3=\mp\theta_2$ and $v_3=v_2$. The mass-squared matrix of the neutral sector factorizes in that case, each factor vanishing linearly with $\lambda_1$. This suggests that these factors are related to the pseudo-Goldstone bosons.

%%%%%%%%%%%%%%%%%%%%%%%%%%%%%%%%%%%%%%%%%%%%%%%%%
\subsubsection{Special cases}
\label{sect:special-cases}
%%%%%%%%%%%%%%%%%%%%%%%%%%%%%%%%%%%%%%%%%%%%%%%%%
As shown in sect.~\ref{sect:neutral-sector}, the mass-squared matrix for the neutral sector has the structure
\begin{equation}
{\cal M}^2_\text{neut}=
\begin{pmatrix}
X & X & X & 0 & 0 \\
X & X & X & 0 & x \\
X & X & X & x & 0 \\
0 & 0 & x & x & x \\
0 & x & 0 & x & x
\end{pmatrix}
\quad
\begin{vmatrix}
\eta_1^\HB \\ \eta_2^\HB \\ \eta_3^\HB \\ \chi_2^\HB \\ \chi_3^\HB
\end{vmatrix}
\end{equation}
where elements that vanish as $\lambda_1\to0$ are denoted by lower-case $x$.
The column to the right is a reminder of the field sequence in the Higgs basis.
If we put $\sin(2\theta_2-2\theta_3)=0$ we get a block-diagonal form with one massless state
\begin{equation}
{\cal M}^2_\text{neut}=
\begin{pmatrix}
X & X & X & 0 & 0 \\
X & X & X & 0 & 0 \\
X & X & X & 0 & 0 \\
0 & 0 & 0 & 0 & 0 \\
0 & 0 & 0 & 0 & x
\end{pmatrix}
\quad
\begin{vmatrix}
\eta_1^\HB \\ \eta_2^\HB \\ \eta_3^\HB \\ \chi_2^\HB \\ \chi_3^\HB
\end{vmatrix}
\end{equation}
The condition $\lambda_1=0$ (instead of $\sin(2\theta_2-2\theta_3)=0$) gives the above texture, only
with a vanishing element on the last row and column ($x\to0$), yielding a block-diagonal form with {\it two} massless CP-odd states.

Finally, for the ``simple model'' of Appendix~\ref{sect:app-minimal} we have
\begin{equation}
{\cal M}^2_\text{neut}=
\begin{pmatrix}
X & X & 0 & 0 & 0 \\
X & X & x & 0 & 0 \\
0 & x & x & 0 & 0 \\
0 & 0 & 0 & x & x \\
0 & 0 & 0 & x & X 
\end{pmatrix}
\quad
\begin{vmatrix}
\eta_1^\HB \\ \eta_2^\HB \\ \chi_3^\HB \\ \chi_2^\HB \\ \eta_3^\HB
\end{vmatrix}
\end{equation}
which is also block diagonal, having interchanged rows (and columns) 3 and 5, i.e., swapped $\eta_3^\HB$ and $\chi_3^\HB$.

%%%%%%%%%%%%%%%%%%%%%%%%%%%%%%%%%%%%%%%%%%%%%%%%%
\section{Gauge couplings}
\label{sect:gauge}
%%%%%%%%%%%%%%%%%%%%%%%%%%%%%%%%%%%%%%%%%%%%%%%%%

The gauge-scalar couplings are determined by the kinetic part of the Lagrangian,
\begin{equation}
{\cal L}_\text{kin}=\sum_{i=1,2,3}(D_\mu\phi_i)^\dagger(D^\mu\phi_i).
\end{equation}

For the cubic gauge-gauge-scalar part, we get
\begin{equation} \label{Eq:VVh}
{\cal L}_{VVh}=\left(gm_WW_\mu^+ W^{\mu-}+\frac{g m_Z}{2\cos\theta_W}Z_\mu Z^\mu\right)\sum_{i=1}^5 O_{i1}h_i,
\end{equation}
with the rotation matrix $O$ relating physical states to the fields of the Higgs basis, as defined by Eq.~(\ref{Eq:Orthogonal}). For the SM-like state at 125.25~GeV, this coupling $O_{i1}$ is severely constrained by the LHC measurements  \cite{ParticleDataGroup:2022pth}. Its magnitude must be close to unity.

For the cubic gauge-scalar-scalar terms, we find
\begin{align} \label{Eq:Vhh}
{\cal L}_{Vhh}&=-\frac{g}{2\cos\theta_W}
\sum_{i=1}^5 
\sum_{j=1}^5(O_{i2}O_{j4}+O_{i3}O_{j5})(h_i\overset\leftrightarrow{\partial_\mu}  h_j)Z^\mu
\nonumber \\
&+\frac{g}{2} \sum_{i=1}^5 \sum_{j=1}^2
[(iO_{i\,j+1}+O_{i\,j+3}) \sum_{k=1}^2 U_{jk}(h_k^+\overset\leftrightarrow{\partial_\mu}  h_i) W^{\mu-} +\text{h.c.}] \nonumber \\
&+\left(ie A^\mu+\frac{ig\cos2\theta_W}{2\cos\theta_W} Z^\mu\right)
\sum_{j=1}^2 (h_j^+\overset\leftrightarrow{\partial_\mu}  h_j^-),
\end{align}
and for the quartic gauge-gauge-scalar-scalar terms, we find
\begin{align}
{\cal L}_{VVhh}&=\biggl(\frac{g^2}{4}W_\mu^+ W^{\mu-} +\frac{g^2}{8\cos^2\theta_W} Z_\mu Z^\mu
\biggr)
\sum_{i=1}^5 h_i^2 \nonumber \\
&+\biggl(\frac{g^2}{2}W_\mu^+ W^{\mu -}
+ e^2A_\mu A^\mu 
+\frac{g^2 \cos^22\theta_W}{\cos^2\theta_W} Z_\mu Z^\mu
+\frac{eg\cos2\theta_W}{\cos\theta_W}A_\mu Z^\mu
\biggr)
 \sum_{j=1}^2 h_j^+ h_j^- \nonumber \\
 &+\biggl[\biggl(
 \frac{eg}{2} W_\mu^+ A^\mu
 -\frac{g^2\sin^2\theta_W}{2\cos\theta_W} W_\mu^+ Z^\mu
 \biggr)
 \sum_{i=1}^5\sum_{j,k=1}^2 U_{jk}h_i h_k^- (O_{ij+1}+iO_{ij+3})+\text{h.c.}
 \biggr].
\end{align}

We have argued that the vicinity of the U(1)$\times$U(1) symmetry should have an impact on the scalar sector, leading to light states that when $\lambda_1\to0$ reveal their Goldstone origin and become odd under CP. In order to shed light on this, we will analyse the coupling of the $Z$ boson to a pair of scalars.
Since $Z$ is odd under CP, it will only couple to the odd component of a two-scalar state $h_i h_j$, not the even part.
This odd component attains its maximal value when one scalar is even and the other is odd. 

A measure of the CP content of two states is obtained from the trilinear coupling $h_ih_jZ$. 
From the first line of equation~(\ref{Eq:Vhh}), an obvious measure is
\begin{equation} \label{eq:P_ij}
P_{ij}=(O_{i2}O_{j4}+O_{i3}O_{j5})-(i\leftrightarrow j).
\end{equation}
We shall refer to it as the ``$Z$ affinity'' of a pair of scalars. A high affinity would mean that the $h_i h_j$ two-scalar state has a significant CP-odd component. Since a two-particle state consisting of two even or two odd scalars would be CP even, we shall somewhat imprecisely refer to the above situation of a large $|P_{ij}|$ as saying the two states have different CP profiles. The quantity $P_{ij}$ is basis-independent, since it refers to a coupling among physical states.

As a reference, it is worth analyzing the $Z$ affinities of pairs of scalars in the CP-conserving 2HDM. We adopt the conventional terminology of $h$ and $H$ being even under CP, whereas $A$ is odd. Furthermore, we take $h$ to be the SM state at 125~GeV. One readily finds that the $Z$ affinity of $h$ and $H$ (both CP even) is zero, whereas that of $H$ and $A$ is unity. However, by the above definition, and in the limit of alignment, the $Z$ affinity of $h$ and $A$ is also zero. With $h=h_j$ aligned, we have $O_{j1}=1$, and (by orthogonality)  $O_{k1}=O_{jk}=0$, with $k\neq j$. Thus, when $h_j$ is aligned, then $P_{kj}=P_{jk}=0$ for all $k$.

Whereas in the 2HDM, allowing for CP violation, the $h_ih_jZ$ couplings are essentially the same as the $h_kZZ$ couplings \cite{Grzadkowski:2014ada}, with $i, j, k$ all different, this is not the case in a 3HDM.

Since $P_{ij}=-P_{ji}$ and $P_{ii}=0$, it follows that there are ten quantities, matching the fact that the rotation matrix $O$ can be generated by ten independent angles. Invoking the orthogonality of the rotation matrix, as well as the 5 independent $h_iVV$ couplings $O_{i1}$, it has been shown that there are in fact only seven independent couplings \cite{Bento:2017eti}.\footnote{This mismatch between the 10 underlying rotation angles and the 7 independent couplings is due to the fact that some sets of rotation angles $(\alpha_{12}, \alpha_{13},\ldots, \alpha_{45})$ and $(\alpha^\prime_{12}, \alpha^\prime_{13},\ldots, \alpha^\prime_{45})$  yield the {\it same} rotation matrix $O$.} We do however find it more transparent to work within this set of ten quantities (\ref{eq:P_ij}), but note from the 2HDM example given above that different CP does not necessarily yield a high value for $|P_{ij}|$. However, a high value for $|P_{ij}|$ can only emerge from states having different CP content.

One may extend the usefulness of the measure of relative CP of two states into the region of small, but non-zero $O_{j1}$ by normalizing it to the squared sum of even and odd couplings,
\begin{equation} \label{eq:P_ij-hat}
\hat P_{ij}=\frac{P_{ij}}{\sqrt{\min(O_{i1}^2,O_{j1}^2)+P_{ij}^2}},
\end{equation}
with $O_{i1}$ representing the CP-even part of the $ZZh_i$ coupling. This measure enhances the ``affinity'' in parameter regions where it would otherwise be small, due to near-alignment.\footnote{This normalization would fail in the zero-measure limit of both $h_i$ and $h_j$ being purely CP odd, i.e., having $\min(O_{i1},O_{j1})=0$ {\it and} $P_{ij}=0$.}

A measure of the CP-odd content of a state can be obtained by summing the square of this coupling over all the other states, $j\neq i$. We denote the square of this quantity $\tilde P_i^2$,
\begin{equation}
\tilde P_i^2=\sum_{j\neq i} P_{ij}^2=\sum_j P_{ij}^2=\sum_{j\neq i} O_{j1}^2=1-O_{i1}^2,
\end{equation}
where in the second step we have used the fact that $P_{ii}=0$, and in the following the orthogonality of $O$.
This has a straightforward interpretation: While we may think of $|O_{i1}|$ as a measure of the CP-even content of $h_i$, we may think of
\begin{equation} \label{Eq:P_tilde}
\tilde P_i=\sqrt{1-O_{i1}^2}
\end{equation}
as the CP-odd part.

%%%%%%%%%%%%%%%%%%%%%%%%%%%%%%%%%%%%%%%%%%%%%%%%%
\section{Yukawa couplings}
\label{sect:Yukawa}
%%%%%%%%%%%%%%%%%%%%%%%%%%%%%%%%%%%%%%%%%%%%%%%%%

With complex vevs, there will also be CP violation in the Yukawa sector, even with real Yukawa couplings. The actual amount of CP violation will depend on how the SU(2) doublets couple to the fermions. As an example, we shall consider natural flavour conservation, where each fermion species couples to at most one Higgs doublet \cite{Glashow:1976nt}. One way to implement this is to let each right-handed fermion sector $u$, $d$ and $e$ couple to a different Higgs doublet according to the following $\ztwo\times \ztwo$ charges
\begin{subequations} \label{eq:yukawa-structure}
\begin{align}
\phi_1 : (+1,+1) & & \phi_2 : (-1,+1) & & \phi_3 : (+1,-1) \\
u_R : (+1,+1) & & d_R : (-1,+1) & & e_R : (+1,-1)
\end{align}
\end{subequations}
Then the Yukawa Lagrangian takes the form 
\begin{align}
\mathcal L_Y &=  \bar Q_L Y^u \tilde\phi_1 u_R +  \bar Q_L Y^d \phi_2 d_R +  \bar E_L Y^e \phi_3 e_R + \text{h.c.}
\end{align}
Expanding the doublets and rewriting the Yukawa neutral interactions in terms of the physical fermion fields, we obtain, in addition to mass terms,
\begin{equation} \label{eq:yukawa}
\mathcal L_Y^\text{neutral} = \frac{1}{v_1}\bar uM^u(\eta_1 +i\chi_1 \gamma_5)u + \frac{1}{v_2}\bar d M^d(\eta_2 +i \chi_2 \gamma_5)d  
+ \frac{1}{v_3}\bar eM^e(\eta_3 + i \chi_3 \gamma_5)e.
\end{equation}
Mixing between the $\eta_i$ and $\chi_i$ fields will cause the neutral physical scalars to have CP violating interactions with the fermions. The Yukawa interaction between a neutral physical scalar $h_i$ and a fermion $f$ takes the  general form
\begin{align}
\mathcal L_{h_iff} = \frac{m_f}{v}h_i(\kappa^{h_iff}\bar f f + i\tilde\kappa^{h_iff} \bar f \gamma_5 f).
\end{align}
This structure can be used to quantify the CP content of the physical scalars. 
For the case of $\tau\bar\tau$ final states, CMS \cite{CMS:2021sdq} has measured this mixing, defined through
\begin{equation}
\tan \alpha^{h_\text{SM}\tau\tau} = \frac{\tilde \kappa^{h_\text{SM}\tau\tau}}{\kappa^{h_\text{SM}\tau\tau}}.
\end{equation}
It has also been suggested to try to measure this quantity for the 2HDM \cite{Fontes:2015mea}.

In order to identify this quantity, we need to express the fields $\eta_i$ and $\chi_i$ of Eq.~(\ref{eq:yukawa}) in terms of the physical scalars, which are not eigenstates of CP. For this purpose, we start by ``undoing'' the transformation to the Higgs basis, Eq.~(\ref{Eq:toHiggsBasis}), writing the inverse, for the neutral fields, in the form
\begin{equation}
\begin{pmatrix}
\eta_1+i\chi_1\\
\eta_2+i\chi_2\\
\eta_3+i\chi_3
\end{pmatrix}
=\tilde{\cal R}^T
\begin{pmatrix}
\eta_1^\HB+iG^0\\
\eta_2^\HB+i\chi_2^\HB\\
\eta_3^\HB+i\chi_3^\HB
\end{pmatrix},
\end{equation}
with $\tilde{\cal R}$ given by Eq.~(\ref{eq:Rtilde}).
Next, the $\eta_i^\HB$ and $\chi_i^\HB$, collectively referred to as $\varphi_i^\HB$ according to Eq.~(\ref{Eq:field-sequence}), can be expressed in terms of the physical states $h_i$ via Eq.~(\ref{Eq:Orthogonal}).

If we introduce a complex quantity for the couplings to $\phi_k$ according to Eq.~(\ref{eq:yukawa}),
\begin{align}
Z_i^{(k)}&=\left(\tilde{\cal R}^T\right)_{k1}O_{i1}+\left(\tilde{\cal R}^T\right)_{k2}(O_{i2}+iO_{i4})
+\left(\tilde{\cal R}^T\right)_{k3}(O_{i3}+iO_{i5}) \nonumber \\
&=\tilde{\cal R}_{1k}O_{i1}+\tilde{\cal R}_{2k}(O_{i2}+iO_{i4})
+\tilde{\cal R}_{3k}(O_{i3}+iO_{i5}),
\end{align}
then for the coupling of $h_i$ to $\tau\bar\tau$ ($k=3$), we have
\begin{equation}
\kappa^{h_i ee} = \frac{v}{v_3} \Re Z_i^{(3)}, \qquad \tilde\kappa^{h_i ee} = \frac{v}{v_3} \Im Z_i^{(3)},
\end{equation}
and
\begin{equation} \label{Eq:Yuk-alpha}
\alpha^{h_i\tau\tau} = \arg(Z_i^{(3)}).
\end{equation}
Some quantitative comments on this quantity will be presented in section~\ref{sect:interpretation}.

As pointed out in the Introduction, this model cannot generate a complex CKM matrix and therefore cannot be considered as the full description.
%%%%%%%%%%%%%%%%%%%%%%%%%%%%%%%%%%%%%%%%%%%%%%%%%
\section{Parameter scans of the scalar potential}
\label{sect:scan}
%%%%%%%%%%%%%%%%%%%%%%%%%%%%%%%%%%%%%%%%%%%%%%%%%
The fact that LHC experiments have determined the Higgs-gauge coupling $h_\text{SM}WW$ to be very close to the SM value shows that the observed Higgs state is essentially pure scalar, with no or very little pseudoscalar admixture. In the notation of Eq.~(\ref{Eq:VVh}), this means that
\begin{equation}\label{E:OSM}
|O_{j1}|\simeq1, \quad \text{for some }j.
\end{equation}

We have performed scans over parameters, analysing the mass spectrum and imposing a condition on the coupling of the SM-like state to two gauge bosons. Each parameter point is required to satisfy boundedness from below, perturbativity and tree-level unitarity. For boundedness from below, only sufficient conditions are known for the $\mathbb{Z}_2\times\mathbb{Z}_2$-symmetric potential \cite{Grzadkowski:2009bt, Faro:2019vcd} and we therefore opt for a numerical check whereas conditions for tree-level unitarity conditions are taken from \cite{Bento:2022vsb}.  
We uniformly sample the parameters in the largest region where all the above constraints can be met\footnote{Alternatively, the scan could be ``factorized'' into a scan over the parameters determining the neutral sector, replacing $\lambda_{ij}$ and $\lambda_{ij}^\prime$ by $\bar\lambda_{ij}$, and another over the charged sector. Qualitatively, the results are found to be similar.}
\begin{subequations}
\begin{alignat}{2}
v_i&\in[0,v],\quad& i&=1,2,3,\quad \text{with\ }v_1^2+v_2^2+v_3^2=v^2, \\
\theta_i&\in[-\pi,\pi], &\quad i&=2,3,\\
\lambda_{ii}&\in[0,4\pi], &\quad i&=1,2,3, \\
\lambda_{ij}, \lambda_{ij}^\prime&\in[-4\pi, 4\pi], &\quad i,j&=1,2,3, \\
\lambda_1&\in[-4\pi,4\pi].
\end{alignat}
\end{subequations}

From these parameters one can reconstruct the mass-squared matrices and diagonalize them.
The neutral mass eigenvalues are ordered as 
\begin{equation}
m_1 < m_2 < m_3 < m_4 <m_5.
\end{equation}

Since the mass-squared matrix is homogeneous in the $\lambda$s, we can rescale the $\lambda$s (all by the same factor) and thereby rescale the masses. The analysis of the sampled parameter points is performed as follows. For each $j=1$ to 5:
\begin{enumerate}
\item
check that the coupling $O_{j1}$ of $h_j$ to $WW$ (or $ZZ$) is compatible with LHC measurements \cite{ParticleDataGroup:2022pth} (at most one value of $j$ will be accepted),
\item
rescale all $\lambda$s such that $m_j=m_\text{SM}=125.25~\text{GeV}$,
\item
apply theoretical cuts (boundedness from below, perturbativity and tree-level unitarity) on all rescaled $\lambda$s (including $\lambda_2$ and $\lambda_3$),
\item
check that the lightest charged scalar is above 80~GeV.
\end{enumerate}
If these conditions are satisfied, the parameter point is kept.
Regarding the LHC measurements of the Higgs-gauge couplings $hVV$ ($V=W,Z$), we use the ATLAS Run 2 value for the coupling modifier $\kappa_V$ \cite{ParticleDataGroup:2022pth} with a $3\sigma$ tolerance, resulting in the following constraint for the SM-like state
\begin{equation}
|O_{j1}| > 0.93.
\end{equation}
Thus we obtain the $h_j$ distribution given in table~\ref{Table:gauge-SM}. 
% A recent ATLAS paper \cite{ATLAS:2021vrm} presents a tighter constraint of $\sigma=0.03$, the distribution for which is also given in the table. 
The theoretical constraints referred to under point 3 are boundedness from below (within the limitation specified above), perturbativity and unitarity.
We note that if these essential experimental and theoretical constraints are to be satisfied then the scenario where $h_1$ is the SM-like state requires fine-tuning of the parameters.

%%%%%%%%%%%%%%%%%%%%%%%%%%%%%%%%%%%%%%%%%%%%%%
\begin{table}[htb]
\caption{Distribution [in \%]  of $h_j$ with gauge coupling $h_j WW$ in agreement with the SM, within $3\sigma$.}
\label{Table:gauge-SM}
\begin{center}
\begin{tabular}{|l|c|c|c|c|c|}
\hline\hline
& $h_1$ & $h_2$ & $h_3$ & $h_4$ & $h_5$ \\
\hline
PDG  \cite{ParticleDataGroup:2022pth} & 0.31 & 38.23 & 28.00 & 22.51 & 10.95 \\
ATLAS \cite{ATLAS:2021vrm} & 0.31 & 38.53 & 27.12 & 21.19 & 12.85 \\
Incl. theoretical cuts & 0.01 & 27.88 & 30.69 & 27.68 & 13.74 \\
\hline\hline
\end{tabular}
\end{center}
\end{table}
%%%%%%%%%%%%%%%%%%%%%%%%%%%%%%%%%%%%%%%%%%%%%

\begin{figure}[htb]
\begin{center}
\includegraphics[scale=0.40]{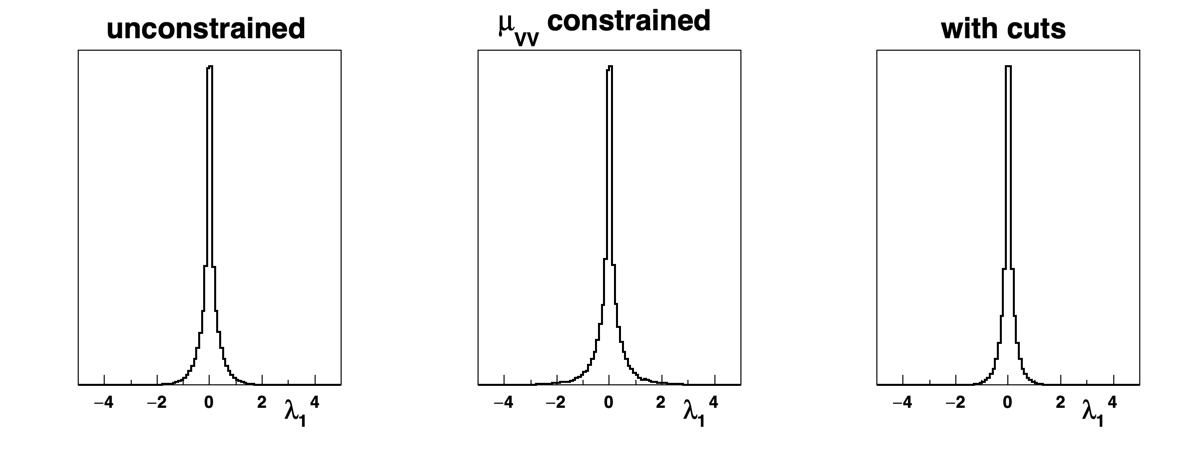}
\end{center}
\vspace*{-4mm}
\caption{Histograms of $\lambda_1$ without (left) and with (center and right) the constraint $|(O_{j1})^2-\kappa_{V}^2|<n\sigma$ with $n=3$. In the right-hand panel we show the impact of imposing further theory constraints (see text).}
\label{Fig:lambda_i}
\end{figure}

We observe that small values of $\lambda_i$ are required to satisfy all the constraints. This is illustrated by fig.~\ref{Fig:lambda_i} where it is seen that the distribution of $\lambda_1$ becomes narrower as the constraint on the $h_iVV$ coupling is applied. The further constraints from boundedness from below and unitarity (right-hand panel) have only a modest impact.
These histograms can be characterized by their r.m.s.\ values:
\begin{equation}
\label{eq:rms-lambda_1}
\lambda_1|_\text{unconstrained}=1.91, \quad
\lambda_1|_{3\sigma}=0.66, \quad
\lambda_1|_{3\sigma+\text{th.~cuts}}=0.37.
\end{equation}
Thus, when the constraint on the $h_i VV$ coupling is imposed, this potential has an approximate U(1)$\times$U(1) symmetry in a sizeable fraction of its viable parameter space.   

%%%%%%%%%%%%%%%%%%%%%%%%%%%%%%%%%%%%%%%%%%%%%%%%

%%%%%%%%%%%%%%%%%%%%%%%%%%%%%%%%%%%%%%%%%%%%%%%%

The parameter points have also been analysed in terms of the average (r.m.s.) $P_{ij}$, representing the coupling of two neutral scalars to the $Z$ boson, defined by Eq.~(\ref{eq:P_ij}). We interpret this as a measure of their relative CP. We have also studied the absolute CP-odd content, as defined by Eq.~(\ref{Eq:P_tilde}). If the average (r.m.s) $P_{ij}$ is large, we say their CP content is different (even if the absolute $\tilde P_i$ and $\tilde P_j$ might be similar), whereas if it is small, we shall say that their CP content is similar.

\begin{figure}[htb]
\begin{center}
\includegraphics[scale=0.30]{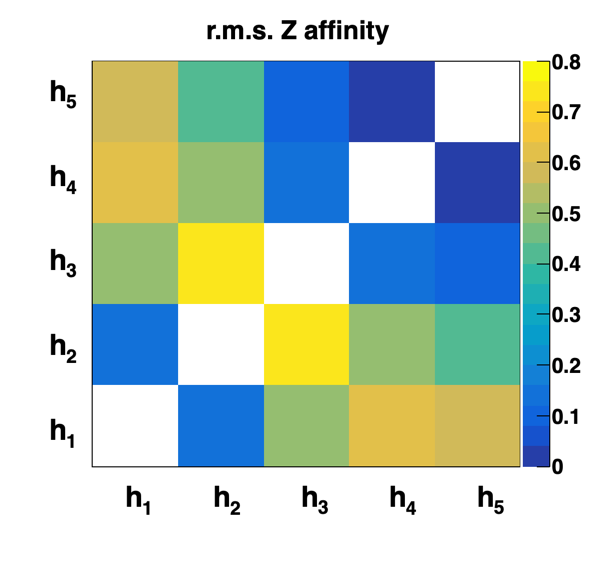}
\includegraphics[scale=0.30]{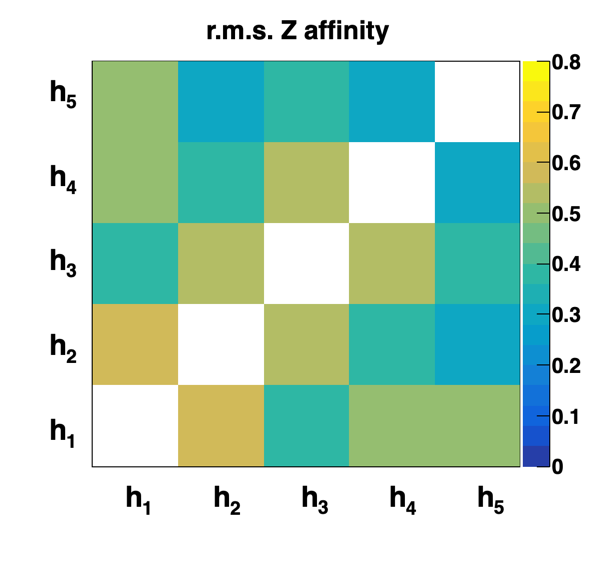}
\end{center}
\vspace*{-4mm}
\caption{Average Z affinity $\Prms{i}{j}$ of states $h_i$ and $h_j$. Left: the U(1)$\times$U(1) limit, as defined by Eq.~(\ref{Eq:low-lambda}); Right: no restriction on the lambdas.}
\label{Fig:affinity}
\end{figure}

For this study, as a reference, we also analysed parameter points that were not subject to the experimental SM-like Higgs constraints described above. 
In Fig.~\ref{Fig:affinity} we compare rms $Z$ affinities for all pairs of neutral scalars, and for two cases, both without the SM-like constraint. In the left panel, we impose a ``near U(1)$\times$U(1) symmetry'' condition
\begin{equation} \label{Eq:low-lambda}
\max(|\lambda_1|, |\lambda_2|, |\lambda_3|)=0.01,
\end{equation}
whereas in the right panel we impose no such constraint, i.e., we do not restrict the scan to the regime of near U(1)$\times$U(1) symmetry. The left panel shows a clear separation into two sets of states, $h_1$ and $h_2$ have low affinity to the $Z$, meaning they have similar CP content, as does the other set, $h_3$, $h_4$ and $h_5$. It is natural to interpret this as follows:
{\it Near the U(1)$\times$U(1) limit we have two neutral states that are approximately odd under CP, and three that are approximately even.} This is fully in accord with the expectations from the Goldstone theorem \cite{Goldstone:1961eq,Goldstone:1962es}, since the Goldstone bosons in the U(1)$\times$U(1) limit will be CP odd \cite{Plantey:2022gwj}.

%%%%%%%%%%%%%%%%%%%%%%%%%%%%%%%%%%%%%%%%%%%%%%%%
%%%%%%%%%%%%%%%%%%%%%%%%%%%%%%%%%%%%%%%%%%%%%%%%

It is instructive to consider how the $Z$ affinity is affected by alignment. Let $h_j$ be ``aligned'', meaning its coupling to $WW$ is maximal, $O_{j1}=1$. By orthogonality, it follows that $O_{k1}=0$ for $k\neq j$ and $O_{jk}=0$, for $k\neq 1$. Then
\begin{equation}
P_{ij}=P_{ji}=0\quad \text{for all }i,
\end{equation}
the aligned scalar $h_j$ has no $Z$ affinity with any other $h_i$ \cite{Plantey:2022gwj}. This is analogous to the CP-even and aligned (and SM-like) $h$ in a CP-conserving 2HDM not having any $Z$ affinity to the pseudoscalar $A$, even though they have opposite CP.

The features displayed in Fig.~\ref{Fig:affinity} change when we turn on the SM-like constraint.
We shall next consider $h_2$ and $h_3$ as candidates for being the discovered state at 125.25~GeV.

%%%%%%%%%%%%%%%%%%%%%%%%%%%%%%%%%%%%%%%%%%%%%%%%%
\section{Accommodating an SM-like state $h_\text{SM}$}
\label{sect:h_SM}
%%%%%%%%%%%%%%%%%%%%%%%%%%%%%%%%%%%%%%%%%%%%%%%%%

Assuming $h_2$ or $h_3$ is identified as $h_\text{SM}$, we shall here first discuss the CP profiles of the light states, as determined from the gauge couplings, and then subsequently study the Yukawa couplings.
%%%%%%%%%%%%%%%%%%%%%%%%%%%%%%%%%%%%%%%%%%%%%%%%%
\subsection{$h_2$ as $h_\text{SM}$}
%%%%%%%%%%%%%%%%%%%%%%%%%%%%%%%%%%%%%%%%%%%%%%%%%

%%%%%%%%%%%%%%%%%%%%%%%%%%%%%%%%%%%%%%%%%%%%%%%%
\begin{figure}[htb]
\begin{center}
\includegraphics[scale=0.30]{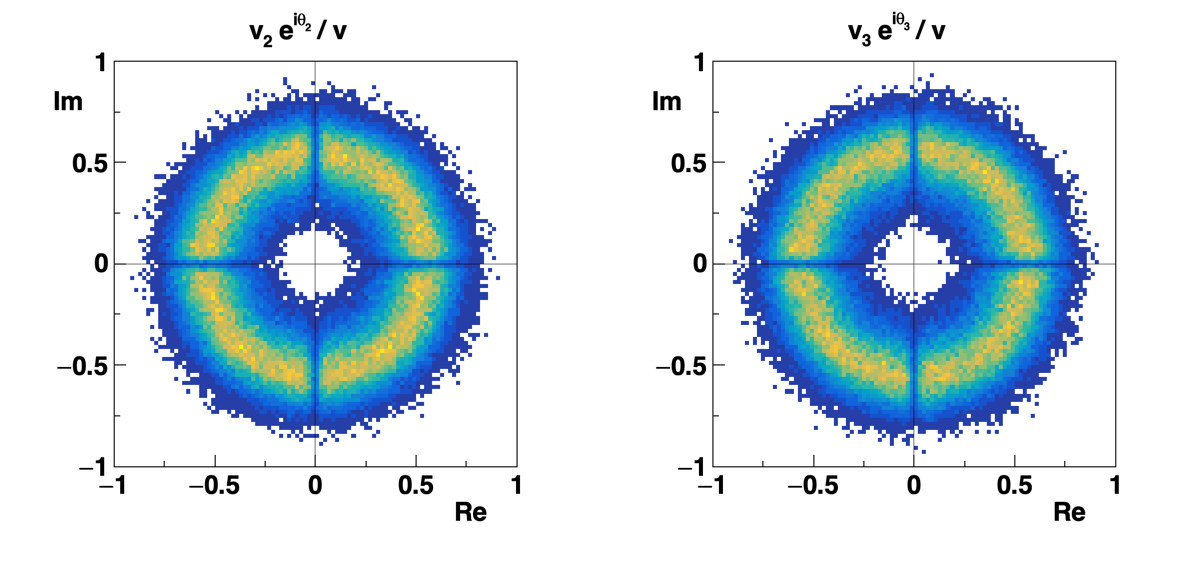}
\end{center}
\vspace*{-4mm}
\caption{Scatter plots of real and imaginary parts of the complex vevs $v_2e^{i\theta_2}/v$ (left) and $v_3e^{i\theta_3}/v$ (right), for $h_2=h_\text{SM}$. The number of surviving parameter points increases when going from dark blue to yellow.}
\label{Fig:v2-v3-2}
\end{figure}
%%%%%%%%%%%%%%%%%%%%%%%%%%%%%%%%%%%%%%%%%%%%%%%%

We first consider the possibility that $h_2$ is to be identified with the discovered SM-like state at 125.25~GeV, as suggested by table~\ref{Table:gauge-SM}.

%%%%%%%%%%%%%%%%%%%%%%%%%%%%%%%%%%%%%%%%%%%%%%%%
\begin{figure}[htb]
\begin{center}
\includegraphics[scale=0.30]{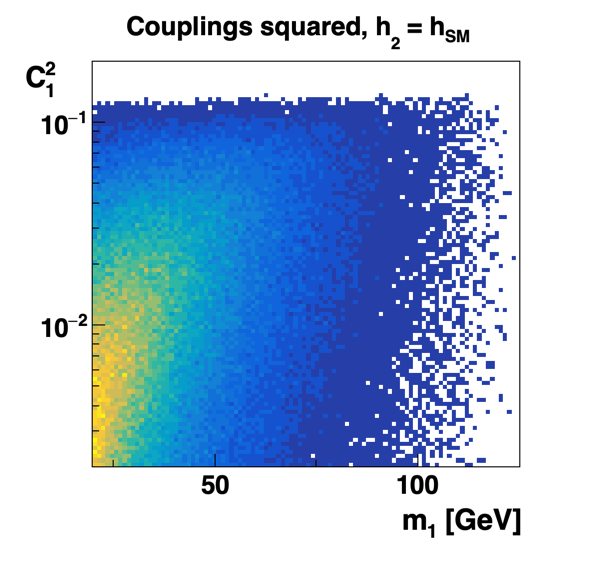}
\end{center}
\vspace*{-4mm}
\caption{Distributions of squared gauge couplings $C_1^2$ of $h_1$ {\it vs} mass (arbitrary units, with yellow ``high'' and dark blue ``low'').}
\label{Fig:couplings-2}
\end{figure}
%%%%%%%%%%%%%%%%%%%%%%%%%%%%%%%%%%%%%%%%%%%%%%%%

%%%%%%%%%%%%%%%%%%%%%%%%%%%%%%%%%%%%%%%%%%%%%%%%
\begin{figure}[htb]
\begin{center}
\includegraphics[scale=0.35]{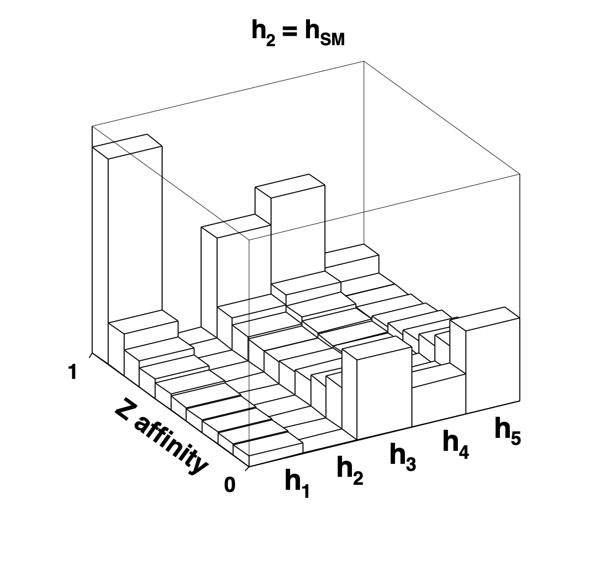}
\end{center}
\vspace*{-12mm}
\caption{Frequency distribution of the relative strength $|\hat P_{2j}|$ of the $h_2h_jZ$ couplings, in units of $g/(2\cos\theta_W)$ (along the $y$-axis) vs $h_j$.}
\label{Fig:HHZ-2}
\end{figure}
%%%%%%%%%%%%%%%%%%%%%%%%%%%%%%%%%%%%%%%%%%%%%%%%

For the parameter points that survive the constraints, we show in figure~\ref{Fig:v2-v3-2} the distributions of the complex vevs $v_2e^{i\theta_2}$ and $v_3e^{i\theta_3}$. Superimposed on circular structures with ``holes'' at $v_2=0$ and $v_3=0$ there are depressions at purely real and purely imaginary values. The latter are due to the fact that $\lambda_2$ and/or $\lambda_3$ become non-perturbative when $|\sin2\theta_2|$ or $|\sin2\theta_3|$ are small.

If $h_2$ were the discovered Higgs particle at 125.25~GeV, why has $h_1$ escaped detection? Searches at LEP \cite{LEPHiggsWorkingGroupforHiggsbosonsearches:2001dnp,McNamara:2002nk} depend on production via the Bjorken mechanism, where the $hZZ$ coupling is essential. But within the present scenario, the $h_1ZZ$ coupling $O_{11}$ is suppressed. This is illustrated in figure~\ref{Fig:couplings-2}, where we plot
\begin{equation}
C_1^2\equiv |O_{11}|^2
\end{equation}
{\it vs} $m_1$. The bulk of the scan points lie at masses below 50~GeV, and for a squared coupling of the order $10^{-2}$. This suppression is simply a result of the unitarity of the mixing matrix $O$.

It is interesting to examine the profile of the neutral state $h_1$ that in this scenario is lighter than 125~GeV. Is it related to the breaking of the U(1) symmetries discussed in the Introduction?
In particular, does it have a significant CP-odd content? Since the gauge field $Z$ is odd under CP, we can ask how large the $h_1 h_2 Z$ coupling is, recalling that in the familiar CP-conserving 2HDM, there is an $HAZ$ coupling of strength 1 [in units of $g/(2\cos\theta_W)$]. The corresponding coupling is for the Weinberg potential given by Eq.~(\ref{eq:P_ij-hat}), from the first term of equation~(\ref{Eq:Vhh}).
We show in figure~\ref{Fig:HHZ-2} the distribution of the $h_2h_jZ$ couplings, in the above units. The strongest coupling is seen to be to $h_j=h_1$, consistent with it having a sizeable CP-odd component.

%%%%%%%%%%%%%%%%%%%%%%%%%%%%%%%%%%%%%%%%%%%%%%%%
\begin{figure}[htb]
\begin{center}
\includegraphics[scale=0.30]{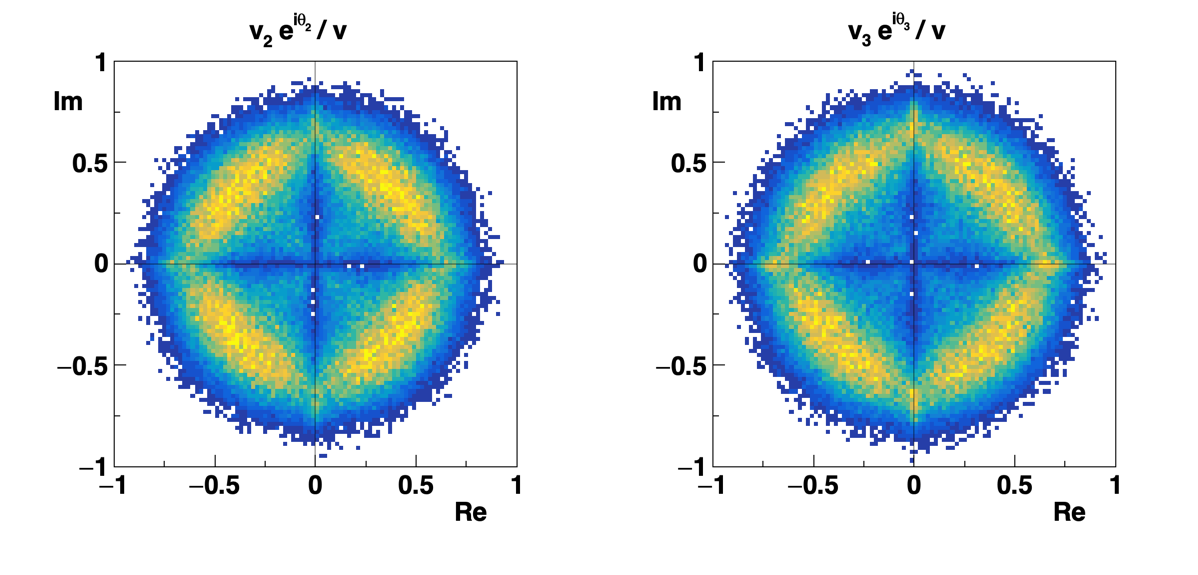}
\end{center}
\vspace*{-4mm}
\caption{Scatter plots of real and imaginary parts of the complex vevs $v_2e^{i\theta_2}/v$ (left) and $v_3e^{i\theta_3}/v$ (right) for $h_3=h_\text{SM}$. Yellow is high, dark blue is low.}
\label{Fig:v2-v3-3}
\end{figure}
%%%%%%%%%%%%%%%%%%%%%%%%%%%%%%%%%%%%%%%%%%%%%%%%

%%%%%%%%%%%%%%%%%%%%%%%%%%%%%%%%%%%%%%%%%%%%%%%%%
\subsection{$h_3$ as $h_\text{SM}$}
%%%%%%%%%%%%%%%%%%%%%%%%%%%%%%%%%%%%%%%%%%%%%%%%%
We next assume that $h_3$ is to be identified as the discovered SM-like scalar. 

For the parameter points that survive the above constraints on maximal allowed value of the $|\lambda|$s and minimum allowed charged Higgs mass, we show in figure~\ref{Fig:v2-v3-3} the distributions of the complex vevs $v_2e^{i\theta_2}$ and $v_3e^{i\theta_3}$. As compared with the previous case, $h_2=h_\text{SM}$, the small-$v_2$ and small-$v_3$ regions are here less depleted.

%%%%%%%%%%%%%%%%%%%%%%%%%%%%%%%%%%%%%%%%%%%%%%%%
\begin{figure}[htb]
\begin{center}
\includegraphics[scale=0.35]{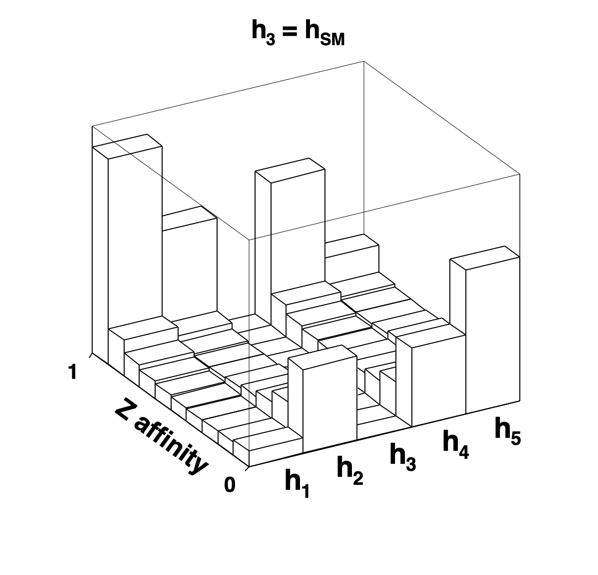}
\end{center}
\vspace*{-12mm}
\caption{Frequency distribution of the relative strength $|\hat P_{3j}|$ of the $h_3h_jZ$ couplings, in units of $g/(2\cos\theta_W)$ (along the $y$-axis) vs $h_j$.}
\label{Fig:HHZ-3}
\end{figure}
%%%%%%%%%%%%%%%%%%%%%%%%%%%%%%%%%%%%%%%%%%%%%%%%

In analogy with the case above, we examine the profile of the neutral states $h_1$ and $h_2$ that in this scenario are lighter than 125~GeV and show in figure~\ref{Fig:HHZ-3} the distribution of $h_3h_jZ$ couplings. The strongest coupling is again seen to be to $h_j=h_1$.

%%%%%%%%%%%%%%%%%%%%%%%%%%%%%%%%%%%%%%%%%%%%%%%%%
\subsection{Yukawa couplings}
\label{sect:interpretation}
\begin{figure}[htb]
\begin{center}
\includegraphics[scale=0.30]{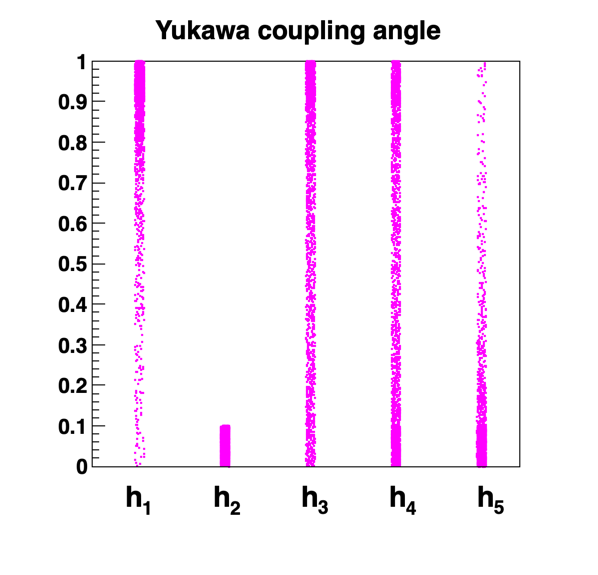}
\includegraphics[scale=0.30]{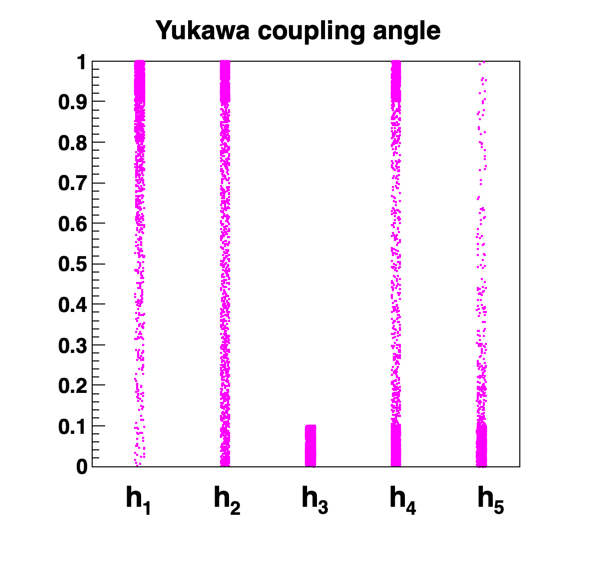}
\end{center}
\vspace*{-4mm}
\caption{Scatter plots of the absolute value of the angle $\alpha$ (in units of $\pi/2$) of Eq.~(\ref{Eq:Yuk-alpha}), characterizing the CP-odd content of the Yukawa couplings to $\tau\bar\tau$ for $h_2=h_\text{SM}$ (left) and $h_3=h_\text{SM}$ (right).}
\label{Fig:Yuk-alpha-narrow}
\end{figure}

%%%%%%%%%%%%%%%%%%%%%%%%%%%%%%%%%%%%%%%%%%%%%%%%%
Returning now to the Yukawa couplings, we study the angle $\alpha$, which is a measure of the relative CP-odd component of this coupling. In Fig.~\ref{Fig:Yuk-alpha-narrow} we show scatter plots\footnote{For better visibility, the points are randomly distributed along the horizontal dimension.} of $\alpha$ (in units of its maximum value, $\pi/2$), for the five different neutral states in the two scenarios $h_2 = h_\text{SM}$ and $h_3 = h_\text{SM}$.
In both cases $h_{\text{SM}}$ is subject to the constraint $|\alpha|<0.1$ which ensures that the CP-odd part of the Yukawa coupling $h_{\text{SM}}\bar\tau\tau$ is consistent with experimental measurements \cite{CMS:2021sdq}. 

This figure supports the feature of the Weinberg potential presented in the introduction: in each scenario, the states lighter than $h_\text{SM}$ are more likely to have a signifi\-cant CP-odd content than the heavier ones.
%%%%%%%%%%%%%%%%%%%%%%%%%%%%%%%%%%%%%%%%%%%%%%%%
%%%%%%%%%%%%%%%%%%%%%%%%%%%%%%%%%%%%%%%%%%%%%%%%

It should be stressed that the results shown in Fig.~\ref{Fig:Yuk-alpha-narrow} depend on how natural flavour conservation is implemented, c.f. (\ref{eq:yukawa-structure}). Because of the symmetry (statistically speaking) of the potential under interchange of $\phi_i$ with $\phi_j$, the scan result does not depend on whether the fermion in question (here, the $\tau$) is coupled to $\phi_1$, $\phi_2$ or $\phi_3$. What is important, though, is the fact that it is coupled to only {\it one} doublet. The outcome would be different if the assumption of natural flavour conservation were relaxed. If $\tau$, e.g., couples to both $\phi_2$ and $\phi_3$, then the angle $\alpha$ would instead be given by
\begin{equation}
\alpha^{h_i\tau\tau}=\arg\left[
\frac{v}{v_2}Z_i^{(2)}+\frac{v}{v_3}Z_i^{(3)}
\right].
\end{equation}

%%%%%%%%%%%%%%%%%%%%%%%%%%%%%%%%%%%%%%%%%%%%%%%%%
\section{Conclusions}
\label{sect:conclusions}
%%%%%%%%%%%%%%%%%%%%%%%%%%%%%%%%%%%%%%%%%%%%%%%%%
 We have explored the spectrum of the Weinberg scalar potential with real coefficients in some detail, determining the CP profiles of the neutral states from how they couple to the electroweak gauge bosons and to fermions. We find that if this potential accommodates the discovered, approximately CP-even Higgs boson at 125.25~GeV, then it naturally (i.e. in the absence of fine-tuning) predicts one or two lighter neutral states. While the model violates CP, one of these states, or both, would have a significant CP-odd content.

One might wonder whether or not imposing the conditions listed in section~\ref{sect:scan} in our parameter scan would bring us close to one of the symmetries obtained for natural alignment in ref.~\cite{Pilaftsis:2016erj}. This would require simple relations among the parameters of the potential \cite{deMedeirosVarzielas:2019rrp,Darvishi:2021txa}.  We have checked that this is not the case. Therefore, the requirement of being close to alignment simply translates into an appropriate choice of parameter space.

In spite of some hints \cite{LEPHiggsWorkingGroupforHiggsbosonsearches:2001dnp,McNamara:2002nk,CMS:2018cyk,Heinemeyer:2021mnz,Biek_tter_2022,Biek_tter_2023}, no state with $m<125~\text{GeV}$ has been observed. This could simply be because in this model the $h_iZZ$ coupling is for the lighter states typically below 10\% of the SM value and production via the Bjorken process is suppressed.

In view of these results, and the appeal of the Weinberg potential, it seems important to pursue the searches for a light scalar, whose coupling to the $Z$ and $W$ is reduced.
In this context, it is important to recall that also branching ratios would differ from those of the SM Higgs. In particular, the $h_j\to\gamma\gamma$ rate would be reduced, again because of the reduced $h_iWW$ coupling and also modified by the loop contributions of the charged scalars. This issue will be discussed elsewhere, the contribution of the charged states could lead to either destructive or constructive interference with the $W$ and fermion loops.

In Ref.\cite{Grzadkowski:2009bt} the same real scalar potential with an additional complex soft symmetry breaking term is studied in a region of parameter space such that the vacuum leaves one of the $\mathbb{Z}_2$ symmetries unbroken, i.e., one of the doublets acquires zero vev. The additional soft term is introduced to explicitly break the two $\mathbb{Z}_2$ symmetries that are also broken by the vacuum. In this way it is possible to have CP violated explicitly by the potential. This framework  results in a viable extension of the Inert Doublet Model \cite{Barbieri:2006dq,Deshpande:1977rw,Cao:2007rm}  providing a good dark matter candidate while having two non-inert doublets.

%%%%%%%%%%%%%%%%%%%%%%%%%%%%%%%%%%%%%%%%%%%%%%%%
\section*{Acknowledgements}
%%%%%%%%%%%%%%%%%%%%%%%%%%%%%%%%%%%%%%%%%%%%%%%%
PO~is supported in part by the Research Council of Norway.
The work of MNR was partially supported by Funda\c c\~ ao 
para  a  Ci\^ encia e a Tecnologia  (FCT, Portugal)  through  the  projects  
CFTP-FCT Unit UIDB/00777/2020 and UIDP/00777/2020, PTDC/FIS-PAR/29436/2017, 
CERN/FIS-PAR/0008/2019, CERN/FIS-PAR/0002/2021, which are  partially  funded  through
POCTI  (FEDER),  COMPETE,  QREN  and  EU. We also thank 
CFTP/IST/University of Lisbon and the University of Bergen, where collaboration visits took place.

%%%%%%%%%%%%%%%%%%%%%%%%%%%%%%%%%%%%%%%%%%%%%%%%%
\appendix

%%%%%%%%%%%%%%%%%%%%%%%%%%%%%%%%%%%%%%%%%%%%%%%%%

%%%%%%%%%%%%%%%%%%%%%%%%%%%%%%%%%%%%%%%%%%%%%%%%%
\section{The mass-squared matrices}
\label{sect:mass-matrix}
%%%%%%%%%%%%%%%%%%%%%%%%%%%%%%%%%%%%%%%%%%%%%%%%%
In this appendix we give the mass-squared matrices of the Weinberg potential.
%%%%%%%%%%%%%%%%%%%%%%%%%%%%%%%%%%%%%%%%%%%%%%%%%
\subsection{Charged sector}
\label{sect:charged-sector}
%%%%%%%%%%%%%%%%%%%%%%%%%%%%%%%%%%%%%%%%%%%%%%%%%
In the charged sector, the elements of the $2\times2$ mass-squared matrix corresponding to the fields $\varphi_{2}^{\text{HB}\, \pm}$ and $\varphi_{3}^{\text{HB}\, \pm}$ can be written as
\begin{subequations}
\begin{align}
({\cal M}^2_\text{ch})_{11}&=-\frac{\lambda_1v^2\sin^2(2\theta_2-2\theta_3)v_2^2v_3^2 }
{\sin2\theta_2\sin2\theta_3 v_1^2 w^2}
-(\lambda^\prime_{12}v_2^2+\lambda^\prime_{13}v_3^2)\frac{v^2}{2w^2}, \\
({\cal M}^2_\text{ch})_{12}&=-\frac{\lambda_1vv_1v_2v_3\sin(2\theta_2-2\theta_3)}{\sin2\theta_2\sin2\theta_3v_1^2w^2}
(v_2^2\sin2\theta_2e^{2i\theta_3}+v_3^2\sin2\theta_3e^{2i\theta_2})
+\frac{vv_1v_2v_3}{2w^2}(\lambda^\prime_{12}-\lambda^\prime_{13}), \\
({\cal M}^2_\text{ch})_{21}&=({\cal M}^2_\text{ch})_{12}^\ast, \\
({\cal M}^2_\text{ch})_{22}&=-\frac{\lambda_1}{\sin2\theta_2\sin2\theta_3w^2}
(2\sin2\theta_2\sin2\theta_3\cos(2\theta_2-2\theta_3)v_2^2v_3^2
+\sin^22\theta_2v_2^4+\sin^22\theta_3v_3^4) \nonumber \\
&-\frac{1}{2w^2}\bigl[(\lambda^\prime_{12}v_3^2+\lambda^\prime_{13}v_2^2)v_1^2
+\lambda^\prime_{23}w^4\bigr].
\end{align}
\end{subequations}
These are all singular if either $\theta_2$ or $\theta_3$ vanishes faster than the other one. The singularities arise due to the constraints (\ref{Eq:lam_23}).

For the rotation to the mass eigenstates $h_{1,2}^+$ we introduce a complex matrix $U$:
\begin{equation}
h_i^+=U_{ij}\varphi_{j+1}^\text{HB +},
\end{equation}
with $\varphi_{2,3}^\text{HB +}$ defined by equation~(\ref{Eq:cap_H_i}). Explicitly, with
\begin{equation}
U=\begin{pmatrix} \cos\gamma & \sin\gamma\, e^{i\phi} \\
-\sin\gamma\, e^{-i\phi} & \cos\gamma
\end{pmatrix},
\end{equation}
we have $h_1^+=\cos\gamma\varphi_2^\text{HB +}+\sin\gamma\, e^{i\phi} \varphi_3^\text{HB +}$ and $h_2^+=-\sin\gamma\, e^{-i\phi} \varphi_2^\text{HB +}+\cos\gamma\varphi_3^\text{HB +}$.

The masses in the charged sector are thus given entirely in terms of $\lambda_1$, $\lambda^\prime_{12}$, $\lambda^\prime_{13}$ and $\lambda^\prime_{23}$, together with the vevs and the phases. The unprimed $\lambda_{ij}$ do not enter. Furthermore, for small $\lambda_1$, either $\lambda^\prime_{12}$ and/or $\lambda^\prime_{13}$ and/or $\lambda^\prime_{23}$ must be negative.

%%%%%%%%%%%%%%%%%%%%%%%%%%%%%%%%%%%%%%%%%%%%%%%%%
\subsection{Neutral sector}
\label{sect:neutral-sector}
%%%%%%%%%%%%%%%%%%%%%%%%%%%%%%%%%%%%%%%%%%%%%%%%%

With the Higgs-basis field sequence (\ref{Eq:field-sequence}), and invoking Eq.~(\ref{Eq:lam_23}), we find
\begin{subequations} \label{Eq:m_sq_gen_neut}
\begin{align}
({\cal M}^2_\text{neut})_{11}&=\frac{4\lambda_1v_2^2v_3^2}{v^2\sin2\theta_2\sin2\theta_3}
[1-\cos(2\theta_2-2\theta_3)\cos2\theta_2\cos2\theta_3] \nonumber \\
&+\frac{2}{v^2}[\lambda_{11}v_1^4+\lambda_{22}v_2^4+\lambda_{33}v_3^4
+\bar\lambda_{12}v_1^2v_2^2
+\bar\lambda_{13}v_1^2v_3^2
+\bar\lambda_{23}v_2^2v_3^2], \label{Eq:m_sq_gen_neut-11}\\
({\cal M}^2_\text{neut})_{12}&=\frac{-2\lambda_1v_2^2v_3^2}{v^2 wv_1\sin2\theta_2\sin2\theta_3}
[\sin^2(2\theta_2-2\theta_3)(2w^2-v^2)-2\cos(2\theta_2-2\theta_3)\sin2\theta_2\sin2\theta_3v_1^2]
 \nonumber \\
&-\frac{v_1}{v^2 w}[2\lambda_{11}v_1^2w^2
-2\lambda_{22}v_2^4-2\lambda_{33}v_3^4
-(\bar\lambda_{12}v_2^2+\bar\lambda_{13}v_3^2)(v^2-2w^2)
-2\bar\lambda_{23}v_2^2v_3^2], \\
({\cal M}^2_\text{neut})_{13}&=\frac{2\lambda_1v_2v_3}{v w\sin2\theta_2\sin2\theta_3}
[v_2^2\sin^22\theta_2-v_3^2\sin^22\theta_3] \nonumber \\
&+\frac{v_2v_3w}{v w^2}[-2\lambda_{22}v_2^2+2\lambda_{33}v_3^2
-\bar\lambda_{12}v_1^2+\bar\lambda_{13}v_1^2+\bar\lambda_{23}(v_2^2-v_3^2)], 
\label{Eq:m_sq_gen_neut-13}\\
({\cal M}^2_\text{neut})_{22}&=\frac{4\lambda_1v_2^2v_3^2}{v^2w^2\sin2\theta_2\sin2\theta_3}
[v_1^2\cos(2\theta_2-2\theta_3)\sin2\theta_2\sin2\theta_3-w^2\sin^2(2\theta_2-2\theta_3)] \nonumber \\
&+\frac{2v_1^2}{v^2w^2}[\lambda_{11}w^4
+\lambda_{22}v_2^4+\lambda_{33}v_3^4
-\bar\lambda_{12} v_2^2w^2-\bar\lambda_{13}v_3^2w^2
+\bar\lambda_{23}v_2^2 v_3^2], \\
({\cal M}^2_\text{neut})_{23}&=\frac{2\lambda_1v_2v_3}{v v_1w^2\sin2\theta_2\sin2\theta_3}
[-w^2\sin(2\theta_2-2\theta_3)(v_2^2\sin2\theta_2\cos2\theta_3+v_3^2\sin2\theta_3\cos2\theta_2) \nonumber \\
&+v_1^2(v_2^2-v_3^2)\cos(2\theta_2-2\theta_3)\sin2\theta_2\sin2\theta_3] \nonumber \\
&+\frac{v_1v_2v_3}{v w^2}[-2\lambda_{22}v_2^2+2\lambda_{33}v_3^2
+(\bar\lambda_{12}-\bar\lambda_{13})w^2+\bar\lambda_{23}(v_2^2-v_3^2)], \\
({\cal M}^2_\text{neut})_{25}&=\frac{2\lambda_1vv_2v_3}{v_1}\sin(2\theta_2-2\theta_3), \\
({\cal M}^2_\text{neut})_{33}&=\frac{-4\lambda_1v_2^2v_3^2}{w^2}\cos(2\theta_2-2\theta_3)
+\frac{2v_2^2v_3^2}{w^2}[\lambda_{22}+\lambda_{33}-\bar\lambda_{23}], \\
({\cal M}^2_\text{neut})_{34}&=\frac{-2\lambda_1vv_2v_3}{v_1}\sin(2\theta_2-2\theta_3), \\
({\cal M}^2_\text{neut})_{44}&=\frac{-2\lambda_1v^2v_2^2v_3^2}{v_1^2w^2\sin2\theta_2\sin2\theta_3}
\sin^2(2\theta_2-2\theta_3), \\
({\cal M}^2_\text{neut})_{45}&=\frac{-2\lambda_1vv_2v_3}{v_1w^2\sin2\theta_2\sin2\theta_3}
\sin(2\theta_2-2\theta_3)
[v_2^2\sin2\theta_2\cos2\theta_3+v_3^2\sin2\theta_3\cos2\theta_2],\\
({\cal M}^2_\text{neut})_{55}&=\frac{-2\lambda_1}{w^2\sin2\theta_2\sin2\theta_3}
[2v_2^2v_3^2\cos(2\theta_2-2\theta_3)\sin2\theta_2\sin2\theta_3
+v_2^4\sin^22\theta_2+v_3^4\sin^22\theta_3],
\end{align}
\end{subequations}
with $({\cal M}^2_\text{neut})_{14}=({\cal M}^2_\text{neut})_{15}=({\cal M}^2_\text{neut})_{24}=({\cal M}^2_\text{neut})_{35}=0$.
Most of these are singular if $\theta_2$ or $\theta_3$ vanishes faster than the other one. 

It is also instructive to study the determinant:
\begin{equation} \label{Eq:det-5by5}
D_{5\times5}=
\frac{\lambda_1^2\sin^2(2\theta_2-2\theta_3)}
{v^2v_1^4(v_2^2+v_3^2)^5\sin^52\theta_2\sin^52\theta_3}
F(\theta_2,\theta_3,\ldots),
\end{equation}
with
\begin{align}
F(\theta_2,\theta_3,\ldots)&=64\lambda_1^3\,v_2^6v_3^{10}w^2\sin^22\theta_2\sin^82\theta_3\tilde F_{2,8}  \nonumber \\
&+\lambda_1^2 v_2^4 v_3^8 \sin^32\theta_2\sin^72\theta_3\, \tilde F_{3,7} \nonumber \\
&+\lambda_1 v_2^2 v_3^6\sin^42\theta_2\sin^62\theta_3\, \tilde F_{4,6} \nonumber \\
&+v_2^4 v_3^4\sin^52\theta_2\sin^52\theta_3\, \tilde F_{5,5} \nonumber \\
&+\{(\theta_2,v_2,\lambda_{22},\bar\lambda_{12})\leftrightarrow(\theta_3,v_3,\lambda_{33},\bar\lambda_{13}))\}
\end{align}
with $\tilde F_{mn}$ regular, homogeneous expansions in the $\lambda$'s and powers of the vevs, as well as sines and cosines of the thetas, accompanying the overall factors $\sin^m2\theta_2\sin^n2\theta_3$. Overall, if both $\theta$'s are small, $F(\theta_2,\theta_3,\ldots)$ is of order ten in the $\theta$'s, cancelling the singularity of the prefactor of Eq.~(\ref{Eq:det-5by5}), but leaving an overall dependence on the thetas given by $\sin^2(2\theta_2-2\theta_3)$.

The determinant of ${\cal M}^2_\text{neut}$ has an overall factor of $\lambda_1^2$ reflecting the fact that in the absence of the terms in $V_\text{ph}$ there would be
 two massless states, originating from the breaking of the U(1)$\times$U(1) symmetry.

For $\sin(2\theta_2-2\theta_3)=0$ the elements $({\cal M}^2_\text{neut})_{25}=({\cal M}^2_\text{neut})_{34}=({\cal M}^2_\text{neut})_{44}=({\cal M}^2_\text{neut})_{45}=0$, and the mass-squared matrix becomes block diagonal. A $3\times3$ block will account for mixing among $\eta_1^\HB$, $\eta_2^\HB$ and $\eta_3^\HB$, whereas a $2\times2$ block will describe a massless $\chi_2^\HB$ and a massive $\chi_3^\HB$. This model would preserve CP, as already mentioned in section~\ref{sect:theta_2=theta_3}. However, there is also another way to achieve factorization, as discussed in Appendix~\ref{sect:app-minimal}.

%%%%%%%%%%%%%%%%%%%%%%%%%%%%%%%%%%%%%%%%%%%%%%%%%
\subsubsection{Masses of the U(1)$\times$U(1) pseudo-Goldstone bosons}
\label{sect:low-lam1}
%%%%%%%%%%%%%%%%%%%%%%%%%%%%%%%%%%%%%%%%%%%%%%%%%

A non-zero $\lambda_1$ explicitly breaks the U(1)$\times$U(1) symmetry of the potential and turns the two Goldstone bosons into pseudo-Goldstone bosons. The masses of these pseudo-Goldstone bosons can be computed to first order in $\lambda_1$ by writing the mass matrix in the symmetry basis as
\begin{align}
{\cal M}_{6\times6}^2 &= {\cal M}_{6\times6}^{2}\Big |_{\lambda_1=0} + \lambda_1 \frac{\partial {\cal M}_{6\times6}^2}{\partial \lambda_1}\\
&\equiv {\cal M}^2_{(0)} + \lambda_1 {\cal M}^2_{(1)},
\label{Eq:perturbedmatrix}
\end{align}
and applying time-independent perturbation theory. The unperturbed system has a threefold degeneracy corresponding to the $\text{U(1)}_Y$ and U(1)$\times$U(1) Goldstone bosons. Hence when $\lambda_1$ is turned on, the $\mathcal O(\lambda_1)$ corrections to the masses of these states are given by the eigenvalues of the perturbation matrix in the degenerate subspace spanned by the three massless states\cite{Sakurai:2011zz}
\begin{gather}
({\cal M}^2_{(1)})_{ij} = \mathbf{n}_i \frac{\partial {\cal M}^2}{\partial \lambda_1} \mathbf{n}^T_j,
\end{gather}
where $\mathbf{n}_i$ ($i=1,2,3$) are three linearly independent massless eigenstates of ${\cal M}^2_{(0)}$. This matrix has a zero eigenvalue due to the fact that the $\text{U(1)}_Y$ Goldstone boson remains massless after $\lambda_1$ is turned on. The two remaining eigenvalues yield the masses of the U(1)$\times$U(1) pseudo-Goldstone bosons at order $\mathcal{O}(\lambda_1)$,
\begin{subequations}
 \begin{equation}
  m^2_{i} = \frac{-\lambda_1}{v_1^2 \sin{2 \theta_2} \sin{2 \theta_3}} 
 \left(v_1^2 v_2^2 \sin^2(2 \theta_2)+v_3^2 v_2^2 \sin^2(2 \theta_2-2 \theta_3)+v_1^2 v_3^2 \sin^2(2 \theta_3) \pm \Delta \right),
 \end{equation}
 where 
  \begin{align}
 \Delta^2 = &\left[v_1^2 \left(v_2^2 \sin^2(2 \theta_2)+v_3^2 \sin^2(2 \theta_3)\right)+v_2^2 v_3^2 \sin^2(2 \theta_2-2 \theta_3)\right]^2 \nonumber \\
 &-4 v_1^2 v_2^2 v_3^2 v^2 \sin^2(2 \theta_2) \sin^2(2 \theta_3)
\sin^2(2 \theta_2-2 \theta_3).
 \end{align}
 \end{subequations}
   Since all masses squared are linear in the lambdas, these above expressions are independent of the lambdas defining $V_0$.

It is instructive to compare these values with the ``simple model'' discussed in Appendix~\ref{sect:app-minimal} for $\theta_3=-\theta_2$ and $v_3=v_2$. In that limit, the above results simplify to
\begin{equation}
m_a^2=4\lambda_1 v_2^2\sin^22\theta_2, \quad
m_b^2=\frac{4\lambda_1v_2^2}{v_1^2}v^2\cos^22\theta_2.
\end{equation}
For a discussion, see Appendix~\ref{sect:app-minimal}.
%%%%%%%%%%%%%%%%%%%%%%%%%%%%%%%%%%%%%%%%%%%%%%%%%

%%%%%%%%%%%%%%%%%%%%%%%%%%%%%%%%%%%%%%%%%%%%%%%%%
\section{A minimal (simple) model}
\label{sect:app-minimal}
%%%%%%%%%%%%%%%%%%%%%%%%%%%%%%%%%%%%%%%%%%%%%%%%%
Inspired by Eq.~(\ref{Eq:constraint-th2-th3}) we see that a minimal version of the model can be constructed by imposing a symmetry under the interchange
\begin{equation}
\phi_2\leftrightarrow\phi_3.
\end{equation}
This immediately implies
\begin{equation}
m_{22}=m_{33}, \quad
\lambda_2=\lambda_3,
\end{equation}
as well as
\begin{equation} \label{Eq:simple}
\lambda_{22}=\lambda_{33}, \quad
\lambda_{12}=\lambda_{13}, \quad
\lambda^\prime_{12}=\lambda^\prime_{13}.
\end{equation}
It follows from the minimization conditions (\ref{Eq:eqs-m_ii}) and (\ref{Eq:angles}) that while the moduli of the vevs are the same, we must have opposite phases,
\begin{equation}
v_2=v_3, \quad \theta_2=-\theta_3.
\end{equation}
Obviously this simple model conserves CP \cite{Branco:1983tn} with CP defined as
\begin{equation} \label{Eq:CP-simple}
\begin{pmatrix}
\langle\phi_1\rangle \\
\langle\phi_2\rangle \\
\langle\phi_3\rangle
\end{pmatrix}
\stackrel{\mbox{CP}}{\longrightarrow} 
\begin{pmatrix}
1 & 0 & 0 \\
0 & 0 & 1 \\
0 & 1 & 0
\end{pmatrix}
\begin{pmatrix}
\langle\phi_1^\ast\rangle \\
\langle\phi_2^\ast\rangle \\
\langle\phi_3^\ast\rangle
\end{pmatrix}.
\end{equation}

Within this framework, the constraints (\ref{Eq:angles}) can be expressed as
\begin{equation} \label{Eq:lambda2}
\lambda_2=-2\lambda_1\frac{v_2^2}{v_1^2}\,\cos(2\theta_2).
\end{equation}

%%%%%%%%%%%%%%%%%%%%%%%%%%%%%%%%%%%%%%%%%%%%%%%%%
\subsection{Charged sector}
%%%%%%%%%%%%%%%%%%%%%%%%%%%%%%%%%%%%%%%%%%%%%%%%%
The mass-squared matrix of the charged sector is found to be given by
\begin{subequations}
\begin{align}
({\cal M}^2_\text{ch})_{11}&=2\lambda_1 \frac{v_2^2}{v_1^2}v^2\cos^22\theta_2-\half \lambda_{12}^\prime v^2 \\
({\cal M}^2_\text{ch})_{12}&=({\cal M}^2_\text{ch})_{21}^\ast=-i\lambda_1 v_2^2\frac{v}{v_1}\sin(4\theta_2), \\
({\cal M}^2_\text{ch})_{22}&=2\lambda_1v_2^2\sin^22\theta_2-\half\lambda_{12}^\prime v_1^2-\lambda_{23}^\prime v_2^2.
\end{align}
\end{subequations}

The two masses are determined by a quadratic equation,
\begin{equation}
m_+^2=\frac{1}{2}[a\pm\sqrt{b}],
\end{equation}
with 
\begin{align}
a&=2\lambda_1\frac{v_2^2}{v_1^2}[v^2-2\sin^2(2\theta_2) v_2^2]
-\lambda_{12}^\prime(v_1^2+v_2^2)-\lambda_{23}^\prime v_2^2, \\
b&=\frac{v_2^4}{v_1^4}
\big\{4\lambda_1^2 (v^2-2\sin^22\theta_2 v_2^2)^2
+4v_1^2(\lambda_{12}^\prime-\lambda_{23}^\prime)[2\sin^22\theta_2(v_1^2+v_2^2)-v^2] 
+(\lambda_{12}^\prime-\lambda_{23}^\prime)^2v_1^4\big\}
\end{align}

If we consider the limit $\lambda_1\to0$, we find
\begin{equation}
m_+^2\to \frac{1}{2}[-\lambda_{12}^\prime(v_1^2+v_2^2\mp v_2^2)-\lambda_{23}^\prime(v_2^2\pm v_2^2)].
\end{equation}

On the other hand, if we make the further assumption that $\lambda_{12}^\prime=\lambda_{23}^\prime$, we find
\begin{subequations}
\begin{align}
m_\alpha^2&=-\frac{1}{2}\lambda_{12}^\prime v^2, \\
m_\beta^2&=m_\alpha^2+\Delta m^2, \\
\Delta m^2&=\frac{2\lambda_1v_2^2}{v_1^2}(v_1^2+2v_2^2\cos^22\theta_3).
\end{align}
\end{subequations}
If $\lambda_1>0$, we have $m_\beta>m_\alpha$, otherwise the order is inverted. We must require $\lambda_{12}^\prime<0$.

%%%%%%%%%%%%%%%%%%%%%%%%%%%%%%%%%%%%%%%%%%%%%%%%%
\subsection{Neutral sector}
%%%%%%%%%%%%%%%%%%%%%%%%%%%%%%%%%%%%%%%%%%%%%%%%%
In the Higgs basis, and invoking Eq.~(\ref{Eq:lambda2}), the $5\times5$ mass-squared matrix takes the form:
\begin{subequations} \label{Eq:mm_neut_no-lam2}
\begin{align}
({\cal M}^2_\text{neut})_{11}&=\frac{2}{v^2}\bigl[-2\lambda_1v_2^4(1+2\cos^22\theta_2)
+\lambda_{11}v_1^4+2\lambda_{22}v_2^4+2\bar\lambda_{12}v_1^2v_2^2+\bar\lambda_{23}v_2^4\bigr], \\
({\cal M}^2_\text{neut})_{12}&=\frac{2\sqrt2\lambda_1v_2^3}{v_1v^2}
(-v_1^2+4v_2^2\cos^22\theta_2) \nonumber \\
&+\frac{\sqrt2 v_1v_2}{v^2}[-2\lambda_{11}v_1^2+2\lambda_{22}v_2^2
+\bar\lambda_{12}(v_1^2-2v_2^2)+\bar\lambda_{23}v_2^2], \\
({\cal M}^2_\text{neut})_{22}&=\frac{v_2^2}{v^2}\bigl[2\lambda_1[-v_1^2
+2(v_1^2+4v_2^2)\cos^22\theta_2]
+(4\lambda_{11}+2\lambda_{22}-4\bar\lambda_{12}+\bar\lambda_{23})v_1^2\bigr], \\
({\cal M}^2_\text{neut})_{25}&=2\lambda_1v \frac{v_2^2}{v_1}\sin4\theta_2,\\
({\cal M}^2_\text{neut})_{33}&=2\lambda_1v_2^2(1-2\cos^22\theta_2)
+(2\lambda_{22}-\bar\lambda_{23})v_2^2, \\
({\cal M}^2_\text{neut})_{34}&=-2\lambda_1v \frac{v_2^2}{v_1}\sin4\theta_2, \\
({\cal M}^2_\text{neut})_{44}&=4\lambda_1v^2\frac{v_2^2}{v_1^2}\cos^22\theta_2,\\
({\cal M}^2_\text{neut})_{55}&=4\lambda_1v_2^2\sin^22\theta_2,
\end{align}
\end{subequations}
the remaining elements being zero.

%%%%%%%%%%%%%%%%%%%%%%%%%%%%%%%%%%%%%%%%%%%%%%%%%
\subsection{Factorization}
%%%%%%%%%%%%%%%%%%%%%%%%%%%%%%%%%%%%%%%%%%%%%%%%%
Since the mass-squared matrix for the neutral sector, Eq.~(\ref{Eq:mm_neut_no-lam2}), becomes block diagonal, its determinant factorizes.
The determinant of this $5\times5$ mass-squared matrix of the neutral sector, equation~(\ref{Eq:mm_neut_no-lam2}), factorizes, the mas-squared matrix becomes block-diagonal.
One factor comes from the $\{\eta_1^\HB,\, \eta_2^\HB,\,\chi_3^\HB\}$ sector,
\begin{equation}
\left({\cal M}^2\right)_{\text{neut},\, 3\times3}
=\begin{pmatrix}
({\cal M}^2_\text{neut})_{11} & ({\cal M}^2_\text{neut})_{12} & ({\cal M}^2_\text{neut})_{15} \\
({\cal M}^2_\text{neut})_{21} & ({\cal M}^2_\text{neut})_{22} & ({\cal M}^2_\text{neut})_{25} \\
({\cal M}^2_\text{neut})_{51} & ({\cal M}^2_\text{neut})_{52} & ({\cal M}^2_\text{neut})_{55}
\end{pmatrix},
\end{equation}
and the other from the $\{\eta_3^\HB,\,\chi_2^\HB\}$ sector,
\begin{equation}
\left({\cal M}^2\right)_{\text{neut},\, 2\times2}
=\begin{pmatrix}
({\cal M}^2_\text{neut})_{33} & ({\cal M}^2_\text{neut})_{34}\\
({\cal M}^2_\text{neut})_{43} & ({\cal M}^2_\text{neut})_{44}
\end{pmatrix}.
\end{equation}
The two determinants are given by
\begin{align} \label{eq:discr-3}
D_{3\times3}=\frac{8\lambda_1v_2^4\sin^2(2\theta_2)}{v_1^2}
&\Bigl[8\lambda_1^2 v_2^4\cos^22\theta_2
-2\lambda_1[\lambda_{11}v_1^4+(4\lambda_{22}+2\bar\lambda_{23})v_2^4\cos^22\theta_2] \nonumber \\
&+v_1^4(2\lambda_{11}\lambda_{22}+\lambda_{11}\bar\lambda_{23}-\bar\lambda_{12}^2)
\Bigr],
\end{align}
and
\begin{equation} \label{eq:discr-2}
D_{2\times2}=\frac{4\lambda_1v^2 v_2^4}{v_1^2}
(-2\lambda_1+2\lambda_{22}-\bar\lambda_{23})\cos^2(2\theta_2).
\end{equation}

Both of these vanish in the limit of $\lambda_i\to0$, i.e., when $V_\text{ph}\to0$. Furthermore, both determinants are proportional to $v_2^4$, so if the vevs of $\phi_2$ and $\phi_3$ were to vanish, two masses in each sector would vanish. This feature is reflected in the scans of the full model shown in figures~\ref{Fig:v2-v3-2} and \ref{Fig:v2-v3-3}. There are no points at the origin in the $v_2\exp(i\theta_2)$ or $v_3\exp(i\theta_3)$ planes.
%%%%%%%%%%%%%%%%%%%%%%%%%%%%%%%%%%%%%%%%%%%%%%%%%
\subsection{CP conservation}
%%%%%%%%%%%%%%%%%%%%%%%%%%%%%%%%%%%%%%%%%%%%%%%%%
Inspection of the gauge couplings discussed in section~\ref{sect:gauge}, in particular the $Zh_ih_j$ couplings given by Eq.~(\ref{eq:P_ij}), shows that $P_{12}=P_{15}=P_{25}=0$ and that $P_{34}=0$, so states within each set have the same CP. Furthermore, the non-vanishing $ZZh_1$, $ZZh_2$ and $ZZh_5$ couplings and the vanishing of the $ZZh_3$ and $ZZh_4$ couplings confirm the following identification:
\begin{alignat}{3}
&\eta^\HB_1, \eta_2^\HB, \chi_3^\HB \text{ (not } \eta_3^\HB)&\quad &\text{mix to form } h_1,\,h_2,\, h_5, &\quad &\text{CP even}, \\
&\chi_2^\HB, \eta_3^\HB \text{ (not } \chi_3^\HB) &\quad &\text{mix to form } h_3,\,h_4, &\quad &\text{ CP odd},
\end{alignat}
and, as stated above, CP is conserved in this model.\footnote{However, because of the mixing between $\eta$ and $\chi$ scalar fields, this model becomes CP violating when coupled to fermions.}

%%%%%%%%%%%%%%%%%%%%%%%%%%%%%%%%%%%%%%%%%%%%%%%%%
\subsection{The two pseudo-Goldstone bosons}
%%%%%%%%%%%%%%%%%%%%%%%%%%%%%%%%%%%%%%%%%%%%%%%%%
In section~\ref{sect:low-lam1} we discussed the masses of the pseudo-Goldstone bosons to first order in $\lambda_1$. For the present simplified model, with $v_3=v_2$ and $\theta_3=-\theta_2$, the results for those masses linear in $\lambda_1$ simplify to
\begin{subequations}
\begin{align}
m^2_i &= \frac{\lambda_1}{v_1^2\sin^22\theta_2}\left[2v_1^2v_2^2\sin^22\theta_2+4v_2^4\sin^22\theta_2\cos^22\theta_2\pm\Delta\right], \\
 \Delta^2  &=4v_2^2\sin^22\theta_2[v_1^2-2(v^2-v_2^2)\cos^22\theta_2]^2.
 \end{align}
\end{subequations}
We find the two values
\begin{equation} \label{eq:m_a-m_b}
m_a^2=4\lambda_1 v_2^2\sin^22\theta_2, \quad
m_b^2=\frac{4\lambda_1v_2^2}{v_1^2}v^2\cos^22\theta_2.
\end{equation}

These mass values are seen to be contained as factors in the above determinants $D_{3\times3}$ and $D_{2\times2}$, with $m_a^2$ being a factor of $D_{3\times3}$ and $m_b^2$ a factor of $D_{2\times2}$. Referring back to the CP properties of the $3\times3$ and the $2\times2$ blocks, we conclude that in the limit $\lambda_1\to0$, then $h_a$ (mass $m_a$) would be even under CP and $h_b$ (mass $m_b$) would be odd.
They become degenerate for
\begin{equation}
|\tan2\theta_2|=\frac{v}{v_1},
\end{equation}
which is necessarily larger than unity.
It follows from the discussion in the previous subsection that such degenerate states would have different CP, as they must.

Finally, in the limit $\lambda_1\to0$ we find compact expressions for the non-pseudo-Goldstone masses: From the $2\times 2$ block
\begin{equation}
m_c^2=(2\lambda_{11}-\bar\lambda_{23})v_2^2,
\end{equation}
and from the $3\times 3$ block
\begin{equation}
m_{d,e}^2 = \frac{\alpha\pm\beta}{2},
\end{equation}
where
\begin{align}
\alpha &= v_2^2 \left(2 \lambda_{22}+\bar\lambda_{23}\right)+2 \lambda_{11} v_1^2, \\
\beta &= \sqrt{4 v_2^2 v_1^2 \left(2\bar\lambda_{12}^2-\lambda_{11} \left(2 \lambda_{22}+\bar\lambda_{23}\right)\right)+v_2^4 \left(2 \lambda_{22}+\bar\lambda_{23}\right)^2+4 \lambda_{11}^2 v_1^4}.
\end{align}

\bibliographystyle{JHEP}

\bibliography{ref}

\end{document}